\documentclass[pre,aps,twocolumn,
pdflatex,
superscriptaddress,floats,floatfix,showpacs,10pt]{revtex4-1}
\usepackage{graphicx}
\usepackage{dcolumn}
\usepackage{bm}
\usepackage{amssymb}
\usepackage{amsmath}
\usepackage{textcomp}
\usepackage{color}
\usepackage{float}

\begin{document}

\title{Radiating Electron Interaction with Multiple Colliding Electromagnetic Waves: Random Walk Trajectories,  L\'evy Flights, Limit Circles, and Attractors\\   (Survey of the Structurally Determinate Patterns)}
\author{S. V. Bulanov$^{1,2}$, T. Zh. Esirkepov$^{1}$, S. S. Bulanov$^{3}$, J. K. Koga$^1$, Z. Gong$^4$, 
X. Q. Yan$^{4,5}$, and M. Kando$^1$\\
$^1${\small {Kansai Photon Science Institute, 
National Institutes for Quantum and Radiological Science and Technology (QST), 
8-1-7 Umemidai, Kizugawa, Kyoto 619-0215, Japan}}\\
$^2${\small {A. M. Prokhorov Institute of General Physics, the Russian Academy of Sciences,\\
Vavilov street 38, 119991 Moscow, Russia}}\\
$^3${\small {Lawrence Berkeley National Laboratory, Berkeley, California 94720, USA}}\\
$^4${\small {State Key Laboratory of Nuclear Physics and Technology,
and Key Laboratory of HEDP of the Ministry of Education,
CAPT, Peking University, Beijing 100871, China}}\\
$^5${\small {Collaborative Innovation Center of Extreme Optics,
Shanxi University, Taiyuan, Shanxi 030006, China}}\\
}

\begin{abstract}
The multiple colliding laser pulse concept formulated in Ref. \cite{SSB-2010a} is beneficial  for achieving an extremely high amplitude of
 coherent electromagnetic field. Since the topology of  electric and magnetic fields oscillating in time 
of multiple colliding laser pulses is far from  trivial and the radiation friction effects are significant in the high field limit,  the dynamics of charged particles interacting with the multiple colliding laser pulses  demonstrates 
remarkable features corresponding to random walk trajectories, limit circles, attractors, regular patterns and L\'evy flights. 
Under extremely high intensity conditions the nonlinear dissipation mechanism  stabilizes the  particle motion 
 resulting in the charged particle trajectory being located within narrow regions and in the occurrence of a new class of regular patterns made by the particle ensembles.
\end{abstract}

\pacs{52.38.-r, 41.60.-m, 52.27.Ep}
\date{\today}
\maketitle

\tableofcontents

\newpage

\section{Introduction}
Recent progress in laser technology has lead to a dramatic increase of laser power and intensity.
The lasers are capable of producing  electromagnetic field intensities well above 
$10^{18}$W/cm$^2$, which corresponds to the relativistic quiver electron energy,
 and in the near future their radiation may reach intensities of 10$^{24}$W/cm$^{2}$ 
and higher \cite{ELIILE}. As a result the laser-matter interaction will happen 
in the radiation friction dominated regimes \cite{MTB, MaShu-2006, ADiP-2012}. 
In a strong electromagnetic field, electrons can be accelerated to such high velocities 
that the radiation reaction starts to play an important role \cite{RAD, ZKS-2002, SVB-2004, RR, RR1, RR2, Thomas_PRX}. 
Moreover,  previously unexplored regimes of the interaction will be entered, in which quantum electrodynamics (QED) 
effects such as vacuum polarization, pair production and cascade development can occur \cite{ADiP-2012, Bell2008}.  

The electromagnetic field intensity of the order of 10$^{24}$W/cm$^{2}$ can be achieved in the focus of a 1$\mu$m wavelength laser 
of ten petawatt power. For 30 fs, i.e. for a ten wave period duration,  the laser pulse energy is about 300 J. 
Within the framework of the multiple colliding laser pulses (MCLP) concept formulated in Ref. \cite{SSB-2010a} 
(see Refs. \cite{SSB-2010b, Gonoskov-2012, Gonoskov-2013, Gelfer-2015} for development of this idea),
the laser radiation with given energy ${\cal E}_{las}$ is subdivided into several beams each of them having $1/N$  of the laser energy,  
where $N$ is the number of the beams. 
If the beams interfere in the focus in a constructive way, i.e. their electric fields are summed, the resulting electric 
field and the laser intensity are equal to $E_N=\sqrt{N} E_{las}$ and to $I_N=N I_{las}$, respectively.
Here $ E_{las}$ and  $ I_{las}$ are the electric field and the intensity of the laser light.  For a large number of  beams 
there is a diffraction constraint on the electric field amplitude in the focus region. In the limit $N\to \infty$ 
the electromagnetic field can be approximated by the 3D dipole configurations (see \cite{SSB-2010b}) 
for which the electric field maximum is given by  \cite{Bassett-1986}
\begin{equation}
E_m=8 \pi \sqrt{\frac{{\cal P}_{las}}{3 c \lambda^2}},
\end{equation}
where ${\cal P}_{las}$, $\lambda$, and $c$ are the laser power, wavelength, and speed of light in vacuum,
respectively.

Since the radiation friction and QED processes both depend  on the particle's momentum, the strength of the present electromagnetic field, 
and on their mutual orientation, it is crucial in understanding the dynamics of charged particles in the electromagnetic 
field in the regime of radiation dominance. Even in the simpliest MCLP case, two counter-propagating plane waves, 
the particle behavior in the standing wave is quite complicated. It demonstrates regular and chaotic motion, random walk, limit circles and strange attractors as is shown by \cite{Mendonca-1983, Bauer-1995, ZMSheng-2002, Lehmann-2012, Lehmann-2016, Gonoskov-2014, Bashinov-2015, Bulanov-2015, Esirkepov-2015, Jirka-2016, Kirk-2016}.
As is well known, the standing wave configuration is widely used in  classical electrodynamics and in  QED theory.  This is due to the fact that in the planes where the magnetic field vanishes, the charged particle may be considered 
interacting with an oscillating pure electric field. This provides  great simplification of the theoretical description. In addition, as has been noted above, in a standing wave formed by two colliding laser pulses, 
the resulting EM field configuration facilitates QED effects (see \cite{SSB-2010a, pairproductiontwopulses, pairproductiontwopulses1}). Computer simulations presented in Refs. \cite{LS-2016,Y-2016,Gonoskov_2016} show that the MCLP concept can be beneficial for realizing such important laser-matter interaction regimes as, for example, the electron-positron pair production 
via the Breit-Wheeler process \cite{LS-2016} and the high efficiency gamma-ray flash generation due to nonlinear Thomson or multi-photon Compton scattering \cite{Y-2016,Gonoskov_2016}. Another configuration for the generation of a gamma-flash is a single laser pulse irradiating an overdense plasma target \cite{R-2012, N-2012, Levy-2016, Corvan-2016}. The applications of the laser based gamma-ray sources are reviewed in Ref. \cite{GALES-2016}. The  radiation friction effects on ion acceleration, on magnetic field self-generation, and  on high-order-harmonics in laser plasmas have been studied in Refs. \cite{Tamburini}, \cite{Macchi}, and \cite{Tang}, respectively.

It is not surprising that the  dynamics of the electron interacting with three-, four-, etc. colliding pulses is even more complicated and rich with novel patterns.

The present paper contains the theoretical analysis of the electron motion in the standing electromagnetic (EM) 
wave generated by two-, three-, and four colliding focused EM pulses. The paper is organized as follows.
In next section we introduce the notations used, describe the field configurations and equations of motion and present the dimensionless  parameters characterizing the charged particle interaction with a high intensity EM 
field. Then, in section 3 we briefly recover the main features of the electron motion in two counter-propagating plane waves. In section 4 we formulate a simple theoretical model of the stabilization of the particle motion in the oscillating field due to  nonlinear dissipation effects, which explains the radiative electron trapping revealed earlier in Refs. \cite{Gonoskov-2012, Gonoskov-2013, Bulanov-2015, Esirkepov-2015, Jirka-2016, Kirk-2016, Ji-2014}. 
Section 5 relates to the regular and chaotic electron motion in three s-polarized laser pulses. 
The radiating electron dynamics in the four s- and p-polarized 
colliding EM pulses is discussed in section 6. Section 7 summarizes the conclusions.

\section{Field configurations,   dimensionless parameters and equations of motion}

\subsection{N colliding  EM waves}

Consider $N$ monochromatic plane waves in vacuum with the same frequencies $\omega_0$ and equal 
amplitudes $a_n$. We assume that the wave vectors ${\bf k}_n$ are in the $(x,y)$ plane. The wave vector of the $n_{th}$ wave is equal to 
\begin{equation}
{\bf k}_n=k_0 [\sin (\theta_n){\bf e}_x+\cos (\theta_n){\bf e}_y],
\end{equation}
 where $k_0=\omega_0/c$, $\theta_n=2\pi (n-1)/N$, $n=1,2,3, ... N$, and 
${\bf e}_x$ and ${\bf e}_y$ are unit vectors in the $x$ and $y$ directions.

It is convenient to describe the s-polarized EM waves with the electric field normal to the $(x,y)$ plane, 
i.e. ${\bf E}=E_z {\bf e}_z$ with the unit vector ${\bf e}_z$ along the $z$ direction, 
in terms of $E_z(x,y,t)$ equal to 
\begin{equation}
E_z
=E_n\sum_{n=1}^N \sin\left \{\omega_0 \left [t-\frac{ \sin (\theta_n)x- \cos (\theta_n) y}{c}\right]\right\}.
\end{equation}
 Here the amplitude of the $n_{th}$ wave is $E_n=E_{0}/\sqrt{N}$ where $E_{0}=E_{las}$. The magnetic field can be expressed 
 by using Maxwell's equations: $(1/c)\partial_t B_x=-\partial_y E_z$ and $(1/c)\partial_t B_y=\partial_x E_z$.
 
 In the case of p-polarized EM waves with the magnetic field normal to the $(x,y)$ plane,
${\bf B}=B_z {\bf e}_z$, the $B_z$ field of colliding $N$ pules is given by 
\begin{equation}
B_z
=B_n\sum_{n=1}^N \cos\left \{\omega_0 \left [t-\frac{ \sin (\theta_n)x- \cos (\theta_n) y}{c}\right]\right\}
\end{equation}
with $B_n=E_{las}/\sqrt{N}$ and the electric field components expressed 
via Maxwell's equations as $(1/c)\partial_t E_x=\partial_y B_z$ and $(1/c)\partial_t E_y=-\partial_x E_z$, respectively.

\subsection{Dimensionless parameters characterizing interaction of laser radiation with charged particles}

Introducing the normalized variables, we change the space and time coordinates to 
$x/\lambda \to x$ and $t \omega/2 \pi \to t$.

The interaction of charged particles with intense EM fields is  characterized  
by several dimensionless and relativistic invariant parameters (\cite{ADiP-2012, Bulanov-2015,Ritus_1}). 

The first parameter is
\begin{equation}
a=\frac{e \sqrt{A_{\mu}A^{\mu}}}{m_e c^2},
\end{equation}
where $A^{\mu}$ is the 4-potential of the electromagnetic field with $\mu =0,1,2,3,4$. Here and below summation over repeating 
indexes is assumed. This parameter is relativistically invariant for a plane EM wave. It is related to the wave normalized amplitude introduced above. 
When it is equal to unity, i.e. the intensity of a linearly polarized EM wave is $I_R=1.37 \times 10^{18}(1\mu m/\lambda)^2{\rm W/cm^2}$,
the quiver electron motion  becomes relativistic. 

The ratio,  $eE/m_e\omega c$, the dimensionless EM field amplitude,  measures 
the work  in units of $m_ec^2$ produced by the field on an electron over the distance equal to the field wavelength. 
Here, $e$ and $m_e$ are the charge and mass of an electron, $E$ and $\omega$ 
are the EM field strength and frequency, and $c$ is the speed of light. 

The second dimensionless parameter is  $\varepsilon_{rad}$:  
\begin{equation}
\label{eq:eps-rad}
\varepsilon_{rad}=\frac{4\pi r_e}{3 \lambda}=1.18\times 10^{-8}\left(\frac{1\mu m}{\lambda}\right),
\end{equation}
which is proportional to the ratio of the classical electron radius $r_e=e^2/m_e c^2=2.8 \times 10^{-13}$cm to the 
laser radiation wavelength, $\lambda$. It essentially determines the strength of the radiation reaction effects for an electron radiating an EM wave.

When one micron wavelength laser intensities exceed $10^{23}$ W/cm$^2$, the nonlinear quantum electrodynamics effects begin  to play a significant role in laser plasma interactions (e.g. see Ref. \cite{Bulanov-2015} and literature cited therein). These effects manifest themselves through multi-photon Compton and Breit-Wheeler effects \cite{Ritus_1,Ritus_2,Ritus_3} (see Refs.: \cite{CBWrecent, CBWrecent1, CBWrecent1bis, CBWrecent2, CBWrecent3, CBWrecent4, CBWrecent5, CBWrecent6, Harvey:2009} for recent studies), \textit{i.e.}, 
through either photon emission by an electron or positron, or electron-positron pair production by a high energy photon, respectively. 
The  multi-photon Compton and Breit-Wheeler  processes are characterized in terms of two dimensionless relativistic and gauge invariant parameters \cite{Ritus_1}: 
\begin{equation}
\label{eq:chieg}
\chi_e=\frac{\sqrt{|F^{\mu \nu} p_{\nu}|^2}}{E_S m_e c } \quad {\rm and} \quad 
\chi_\gamma = \frac{\lambda_C\sqrt{|F^{\mu\nu}k_\nu|^2}}{E_S}.
\end{equation}
where $p_\nu$ and $\hbar k_\nu$ denote  the 4-momenta of an electron or positron undergoing the Compton process and a photon undergoing the Breit-Wheeler process, 
the 4-tensor of the electromagnetic field is defined as $F_{\mu \nu}=\partial_{\mu} A_{\nu}-\partial_{\nu} A_{\mu}$, with the critical QED electric field
\begin{equation}
\label{eq:Es}
E_S=\frac{m_e^2c^3}{e\hbar}.
\end{equation}
This field is also known as the ``Schwinger field'' \cite{BLP-QED}. Its amplitude is about $10^{18}$V/cm, which corresponds to the radiation intensity  
$\approx 10^{29}$ W/cm$^2$. The work  produced by the field $E_S$ on an electron over the distance equal to the reduced Compton wavelength, 
$\lambda_C=\hbar/m_ec=3.86 \times 10^{-11}cm$  equals $m_ec^2$. Here $\hbar$ is the reduced Planck constant. 

In 3D notation the parameter $\chi_e$ given by Eq. (\ref{eq:chieg}) reads
	\begin{equation}
	 \chi_e=\frac{\gamma_e}{E_S}\sqrt{\left({\bf E}+\frac{{\bf p}_e\times {\bf B}}{m_e c \gamma_e}\right)^2
	 -\left(\frac{{\bf p}_e\cdot {\bf E}}{m_e c \gamma_e}\right)^2}.
	\label{eq:chi-e-3D}
	\end{equation}
For the parameter $\chi_{\gamma}$ defined by Eq. (\ref{eq:chieg}) we have
	\begin{equation}
	 \chi_{\gamma}=\frac{\hbar}{E_Sm_ec}\sqrt{\left(\frac{\omega_{\gamma}}{c}{\bf E}+{\bf k}_{\gamma}\times {\bf B}\right)^2
	 -\left({\bf k}_{\gamma}\cdot {\bf E}\right)^2}.
	\label{eq:chi-g-3D}
	\end{equation}
	Here $\gamma_e$, ${\bf p}_e$, $\omega_{\gamma}$ and ${\bf k}_{\gamma}$ correspond to the representation of the 
	electron 4-momentum $p_{\nu}$ and of the photon 4-wavenumber 
	$k_{\nu}$ as $p_{\nu}=(\gamma_e m_ec, {\bf p})$ and $k_{\nu}=(\omega_{\gamma}/c, {\bf k}_{\gamma})$, respectively. The parameter $\chi_e$ can also be defined as the ratio of the electric field to the critical electric field of quantum electrodynamics, $E_S$, in the electron rest frame. In particular, it characterizes the probability of the gamma-photon emission by an electron with 4-momentum $p_{\nu}$ in the field of the electromagnetic wave, in the Compton scattering process. 

The parameter $\chi_\gamma$ characterizes the probability of the electron-positron pair creation by the photon  
with the momentum $\hbar k_\nu$ interacting with a strong EM wave in the Breit-Wheeler process.

 The probabilities of the Compton scattering and of  the Breit-Wheeler processes depend strongly on $\chi_e$ and $\chi_\gamma$,
  reaching optimal values when $\chi_e\sim 1$ and $\chi_\gamma\sim 1$ (\cite{Ritus_1}). 
  
  In the case of an electron interaction with a plane EM wave propagating along the $x$-axis with phase and group velocity equal to speed of light in vacuum 
  the parameters of the interaction can be written in terms of EM field strength, normalized by the QED critical field given by Eq. (\ref{eq:Es}),  
  and either the electron $\gamma_e$-factor or the photon energy $\hbar \omega_{\gamma}$:  
	\begin{equation}
	\label{eq:chiecop}
	\chi_e=\frac{E}{E_S}\left(\gamma_e-\frac{p_x}{m_ec}\right)
	\end{equation}
	 and 
	\begin{equation}
	 \chi_\gamma=\frac{E}{E_S}\frac{\hbar(\omega_{\gamma}-k_{\gamma, x} c)}{m_e c^2}.
	 \end{equation}
  For an electron interacting with the EM wave the linear combination of the electron energy and momentum, 
	\begin{equation}
	h_e=\gamma_e-p_x/m_ec,
	 \end{equation}
	 on r.h.s. of Eq. (\ref{eq:chiecop}) is an integral of motion (\cite{LLTF}). Its value 
  is determined by initial conditions.
  
  If an electron/positron or a photon co-propagates with the EM wave, then in the former case the parameter $\chi_e$ is suppressed by a factor $(2\gamma_{e,0})^{-1}$, 
  i.e $\chi_e\simeq(2\gamma_{e,0})^{-1}(E/E_S)$, where $\gamma_{e,0}$ is the electron gamma-factor before interaction with the laser pulse. 
  In the later case, when the gamma-photon co-propagates with the EM wave, the parameter $\chi_\gamma$ is equal 
  to zero, $\chi_\gamma=0$, because $\omega_{\gamma}=k_{\gamma, x} c$. On the contrary, the parameter $\chi_e$ can be enhanced to approximately $2\gamma_{e,0} E/E_S$, 
  when the electron interacts with a counter-propagating laser pulse.  
  Therefore the head-on collision configuration has an apparent advantage for strengthening the electron-EM-wave 
  interaction and, in particular, for enhancing the $\gamma$ ray production due to 
  nonlinear Thomson or/and Compton scattering.

 Depending on the energy of charged particles 
and field strength the interaction happens in one of the following regimes parametrized by the values of $a$, $\chi_{e}$, and  $\chi_{\gamma}$:

 (i) $a>1$, the relativistic interaction regime (\cite{MTB}), 
 
 (ii) $a>\varepsilon_{rad}^{-1/3}$, the interaction becomes radiation dominated (\cite{ZKS-2002, SVB-2004, BK-2013}), 
 
 (iii) $\chi_e \geq 1$ the quantum effects begin to manifest themselves (\cite{diPiazza-2010, NIMA-2011, Bulanov-2015}), and 
 
 (iv) $\chi_e>1$, $\chi_\gamma>1$ marks the condition for the EM avalanche (\cite{SSB-2010b, Fedotov-2010, Elkina-2011, Nerush-2011, SSB-2013}),  
 which is the phenomenon of exponential growth of the number of electron-positrons and photons in the strong EM field,  being able to develop. These conditions can be supplemented by $\alpha a>1$, which indicates that the number of photons emitted incoherently per
laser period can be larger than unity as has been noted by \cite{diPiazza-2010}.
Here the parameter $\varepsilon_{rad}$ is given by Eq. (\ref{eq:eps-rad}) and $\alpha=e^2/\hbar c\approx 1/137$ is the fine structure constant.

\begin{figure*}
\begin{center}
\includegraphics[keepaspectratio=true,width=12 cm]{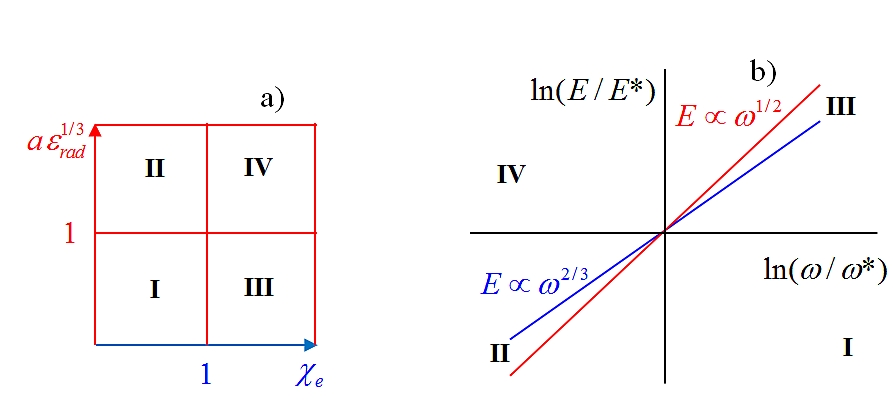}
\end{center}
\caption{ Regimes of electromagnetic field interaction with matter on the plane of parameters: 
a) the normalized EM wave amplitude $a \varepsilon^{1/3}_{rad}$ and the parameter $\chi_e$; 
b) accordingly  the $(\ln(E/E^*),\ln(\omega/\omega^*))$ plane, where $E^*$ and $\omega^*$ are given by 
Eqs. (\ref{eq:Elstar}) and (\ref{eq:omegastar}), respectively. 
The parameter planes are subdivided into 4 domains:
(I) Electron - EM field interaction in the particle dominated radiation reaction domain;
(II) Electron - EM field interaction is dominated by the radiation reaction; 
(III) Electron - EM field interaction is in the particle dominated QED regime;
(IV) Electron - EM field interaction is in the radiation dominated QED regime.
}
\label{Fig:a-chi_e}
\end{figure*}

As one can see two dimensionless parameters, $a$ and $\chi_e$,  can be used to subdivide the $(a,\chi_e)$ plane into four domains shown in Fig. \ref{Fig:a-chi_e} a) (see also Refs. \cite{Bulanov-2015, SVB-PPCF-2016}). 
The $\chi_e=1$ line divides the plane into the radiation reaction description of the interaction domain ($\chi_e<1$) and QED description of interaction domain ($\chi_e>1$). The $a=\varepsilon_{rad}^{-1/3}$ line divides the plane into radiation dominated ($a>\varepsilon_{rad}^{-1/3}$) and particle dominated ($a<\varepsilon_{rad}^{-1/3}$) regimes of interaction domains. We note that the $a=\varepsilon_{rad}^{-1/3}$ threshold comes from the requirement for an electron to emit the amount of energy per EM wave period equal to the energy gain from the EM wave during the wave period. If one takes into account the discrete nature of the photon emission, then the same condition will take the form $a\, m_e c^2=\hbar \omega_{\gamma} (\lambda/L_R)$ \cite{Ritus_3},where $L_R$ is the radiation length. It is of the order of \cite{Bolotovskii-1966}
\begin{equation}
\label{eq:astar}
L_R=\lambda_{\gamma} \gamma_e^2. 
\end{equation}
In the limit $\chi_e<<1$, when $\lambda_{\gamma}\approx \lambda/\gamma_e^3$  and $\gamma_e\approx a$ we have 
$L_R\approx2\lambda/a$. For $\chi_e>>1$  the radiation length is $L_R\approx\lambda \gamma_e^{1/3}/a^{2/3}$ as shown in Refs. \cite{Ritus_1,Ritus_2,Ritus_3}. This condition in the limit $\chi_e\rightarrow 0$ tends to the classical limit $a=\varepsilon_{rad}^{-1/3}$.   
 
The intersection point, where $a_{rad}=\varepsilon_{rad}^{-1/3}$ and the parameter $\chi_e$ is equal to unity, determines critical values of the EM wave amplitude $\varkappa_a  a^*$ with
\begin{equation}
\label{eq:astar}
a^*=\left(\frac{3 c}{2 r_e \omega^*}\right)^{1/3}=\frac{ \hbar c}{e^2}=\frac{1}{\alpha}, 
\end{equation}
i. e. the wave electric field is $\varkappa_a \varkappa_{\omega} E^*$, where
\begin{equation}
\label{eq:Elstar}
E^*=E_S \alpha,
\end{equation}
and the wave frequency $\varkappa_{\omega} \omega^*$ with $\omega^*$ given by
\begin{equation}
\label{eq:omegastar}
\omega^{*}=\frac{e^4 m_e}{\hbar^3}=\frac{m_e c^2}{\hbar \alpha^2}. 
\end{equation}
Here $\alpha=1/137$ is the fine structure constant, and $\varkappa_a$ and $\varkappa_{\omega}$ are constants 
of the order of unity. The normalized EM wave amplitude equals $a^*=137 $ 
with corresponding the wave intensity $I^*=2.6 \times 10^{22}$W/cm$^{2}$.  The corresponding photon energy 
is $\hbar \omega^{*}=m_e c^2/ \alpha^2 \approx 27\,$eV. We note that the value of $a^*=1/\alpha$ corresponds to the one of conditions for the charged particle interaction with EM field to be in the QED regime, $\alpha a>1$ (see also \cite{diPiazza-2010}).

Concrete values of the coefficients  $\varkappa_a$ and $\varkappa_{\omega}$ depend on the specific electromagnetic 
configuration. For example, in the case of a rotating homogeneous electric field (it can be formed in 
the antinodes of an electric field in the standing EM wave) analyzed in Ref. \cite{Bulanov-2015}, they are  $\varkappa_a=3$ and 
$\varkappa_{\omega}=1/18$, respectively, which gives $\varkappa_{a}a^*=411 $, with the intensity equal to 
$ 2.3 \times10^{23}$W/cm$^{2}$, and $\varkappa_{\omega} \hbar \omega^{*}=m_e c^2 \alpha^2/18\approx 1.5\,$eV.

Here we would like to attract attention to the relationship between the well known critical electric field of classical 
electrodynamics $E_{cr}$, the  critical electric field of quantum 
electrodynamics $E_{S}$ and the electric field $E^*$. They can be written as $E_{cr}=e/r_e^2$, 
$E_{S}=e/r_e \lambda_C$, and $E^*\approx e/ \lambda_C^2$, respectively. In other words we have 
$E_{S}=E_{cr}\alpha$, and $E^*=E_{cr}\alpha^2$.

Using the relationships obtained above we find that on the line $a \varepsilon^{1/3}_{rad}=1$ the wave 
electric field is proportional to the frequency in the $2/3$ power, i. e. $E/E^*=(\omega/\omega^*)^{2/3}$, 
and on the line $\chi_e=1$ we have $E/E^*=(\omega/\omega^*)^{1/2}$.

Fig. \ref{Fig:a-chi_e} b) shows the $(\ln (E/E^*), \ln(\omega/\omega^*))$ plane with 4 domains.
The lines intersect each other at the point $(0,0)$, i.e. at the point where $E=E^*$ and $\omega=\omega^*$.

\subsection{Radiation friction force with the QED form-factor}

	 In order to describe the relativistic electron dynamics in the electromagnetic field we shall use 
	 the equations of electron motion: 
\begin{equation}
\frac{d{\bf p}}{d t}=e\left({\bf E}+\frac{\bf v}{c} \times {\bf B} \right)+ F_{rad},
\label{eq:equmot-mom}
\end{equation}
\begin{equation}
\frac {d{\bf x}}{d t}=\frac{\bf p}{m_e  \gamma},
\label{eq:equmot-coord}
\end{equation}
where  the radiation friction force, $F_{rad}=G_e {\bf f}_{rad}$, is the product of the classical radiation friction force, ${\bf f}_{rad}$, in the Landau-Lifshitz form (\cite{LLTF}): 	 
\[ {\bf f}_{rad}=\frac{2 e^3}{3 m_e c^3 \gamma} 
\left\{ \left(
\partial_t +({\bf v}\nabla 
\right){\bf E}
+\frac{1}{c} 
\left[
{\bf v} \times  \left(\partial_t +({\bf v}\nabla ){\bf B}\right]\right) \right \} \] 
\begin{equation}
+\frac{2 e^4}{3 m_e^2 c^4} \left \{{\bf E} \times {\bf B}+\frac{1}{c} \left[ {\bf B}\times \left({\bf B}\times {\bf v}\right)+{\bf E}\left( {\bf v}\cdot {\bf E}\right)\right]
\right\}
\label{eq:equmot-frad-LL}
\end{equation}
\[-\frac{2 e^4}{3 m_e^2 c^5 } \gamma^2  {\bf v}\left \{ \left({\bf E}+ \frac{1}{c}{\bf v} \times {\bf B}\right)^2 - \frac{1}{c^2}\left( {\bf v}\cdot {\bf E}\right)^2 \right\}
\]
and a form-factor $ G_e $, which takes into account the quantum electrodynamics weakening of the radiation friction \cite{BLP-QED, RAD-QED, RAD-QED1, RAD-QED2, RAD-QED3}. Discussions of the relationship between the Landau-Lifshitz and Lorentz-Abraham-Dirac forms of the radiation friction force and what form of the force follows from the QED calculation, can be found in Refs. \cite{LL-LAD, ZHANG, IldertonRRQED}
 and in the literature cited therein.
	 
	As we have noted above, the threshold of the QED effects is determined by the dimensionless parameter $\chi_e$ given by Eq. (\ref{eq:chi-e-3D}). For example, if  an electron moves in the magnetic field $B$, the parameter is equal to $\chi_e\approx \gamma_e (B/B_S)$, where $B_S=m_e^2c^3/e\hbar$ 
 is the QED critical magnetic field (see also Eq. (\ref{eq:Es})). 
 The energy of the emitted synchrotron photons is 
\begin{equation}
\hbar \omega_{\gamma}=m_e c^2 \gamma_e\frac{ \chi_e}{2/3+\chi_e}.
\label{eq:omQED}
\end{equation}
In the limit $\chi_e\ll 1$ the frequency $\omega_{\gamma}$ is equal to $(3/2)\omega_{Be}\gamma_e^2$ in accordance with classical electrodynamics (see \cite{LLTF}). Here $\omega_{Be}=e B/m_e c$ is the Larmor frequency.
If $\chi_e\gg 1$ the photon energy is equal to the energy of the radiating electron: $\hbar \omega_{\gamma}=m_e c^2 \gamma_e$.

The radiation friction force in the limit $\gamma_e \to \infty$, i.e. the last term on the r.h.s. of Eq. (\ref{eq:equmot-frad-LL}) retained, can be written in the following form (see also Refs. \cite{Bulanov-2015, RAD-QED, RAD-QED1, RAD-QED2, RAD-QED3} and literature cited therein)
\begin{equation}
\label{eq:fradQ}
{\bf f}_{rad}=-\frac{2 \, \alpha\, c \, G_e(\chi_e)\, \chi_e^2}{3\lambda_C }{\bf p}.
\end{equation}
Here the QED effects are incorporated into the equations of the electron motion by using the form-factor $G_e(\chi_e)$ (see Ref. \cite{RAD-QED}), which is equal to the ratio of full radiation intensity to the intensity of the radiation emitted by a classical electron. It reads 
\begin{equation}
G_e(\chi_e)=
\frac{3}{4}\int^{\infty}_0\left[\frac{4+5\chi_ex^{3/2}+4\chi_e^2x^{3}}{\left(1+\chi_ex^{3/2}\right)^4} \right]
\Phi^{\prime}(x)xdx,
\label{eq:GeChi}
\end{equation}
where $\Phi(x)$ is the Airy function (\cite{AS}). In Eq. (\ref{eq:fradQ}) we neglect the effects of the discrete nature of the photon emission in quantum electrodynamics (see \cite{Thomas_PRX, Bashinov-2015, SSB-2013, Esirkepov-2015, Jirka-2016, DUCLOUS, BRADY}).
	
	In the limit $\chi_e\ll1$ the form-factor $G(\chi_e)$
	tends to unity as 
\begin{eqnarray}
	G_e(\chi_e)=1-\frac{55 \sqrt{3}}{16}\chi_e+48 \chi_e^2
	+ ... 
	\\ \nonumber
	\approx 1-5.95 \chi_e+48 \chi_e^2
	+...\,.
	\label{G-chi-e0}
	\end{eqnarray}
	For $\chi_e\gg1$ it tends to zero as
	\begin{eqnarray}
	G_e(\chi_e)= \frac{32 \pi}{27\, 3^{5/6} \Gamma (1/3)\chi_e^{4/3}}-\frac{1}{\chi_e^{2}}+... 
	\\ \nonumber
	\approx \frac{0.5564}{ \chi_e^{4/3}}-\frac{1}{\chi_e^{2}}
	+...\,.
	\label{G-chi-eInf}
	\end{eqnarray}

	However expression (\ref{eq:GeChi}) and the asymptotical dependences (\ref{G-chi-e0}) and (\ref{G-chi-eInf}) 
	are not convenient for implementing them in the computer codes.
    For the sake of calculation simplicity  we shall use the following approximation 	
	\begin{equation}
	G_{R}(\chi_e)\approx \frac{1}{\left(1+8.93\chi_e + 2.41\chi_e^2\right)^{2/3}}.
	\label{G-chi-eappr}
	\end{equation}
	Within the interval $0<\chi_e<10$ the accuracy of this approximation is better than 1$\%$.

\section{Electron motion in the standing EM wave formed by two counter-propagating EM pulses}

\subsection{EM field configuration}

An electron interaction with an EM field formed by two counter-propagating waves was addressed a number of times in high field theory using classical quantum electrodynamics approaches because it provides one of the basic EM configurations where important properties of a radiating electron can be revealed  (e.g. see above cited publications  \cite{ADiP-2012, Gonoskov-2013, Mendonca-1983, Lehmann-2012, Lehmann-2016, Gonoskov-2014,  Bashinov-2015, Bulanov-2015, Esirkepov-2015, Lobet-2015, Chang-2015, Jirka-2016, Kirk-2016, BaKuKi-2016, Grismayer-2016}).   Here we present the results of the analysis of an electron motion in a standing EM wave in order to compare them below with the radiating electron behavior in a more complicated EM configuration formed by three and four waves with various polarizations.
  
Here we consider an electron interaction with the electromagnetic field corresponding to two counter-propagating linearly polarized waves of equal amplitudes, $(a_{0}/{2}) \cos(t+x)$ and $(a_{0}/{2}) \cos(t-x)$, forming a standing wave. The field is given by the 
	 electromagnetic 4-potential
	  \begin{equation}
{\bf A}=a_{0}\cos{t}\cos{x}\, {\bf e}_z.
\label{eq:4-poten2waves}
\end{equation}
This is a standing electromagnetic wave with zero magnetic and electric field nodes located at the coordinates $x=\pm \pi n$ and $x=\pm \pi (n+1/2)$ with $n=0,1,2, ...$, respectively.
	 
	 Numerical integration of the electron motion equations  
	 with the radiation friction force in the form (\ref{eq:fradQ}) shows different features of the electron 
	 dynamics depending on the electromagnetic wave amplitude and the dissipation parameter $\varepsilon_{rad}$.
	 
	 \subsection{Relatively weak intensity limit}
	 
\begin{figure*}
	   \begin{center}
   \includegraphics[width=12 cm]{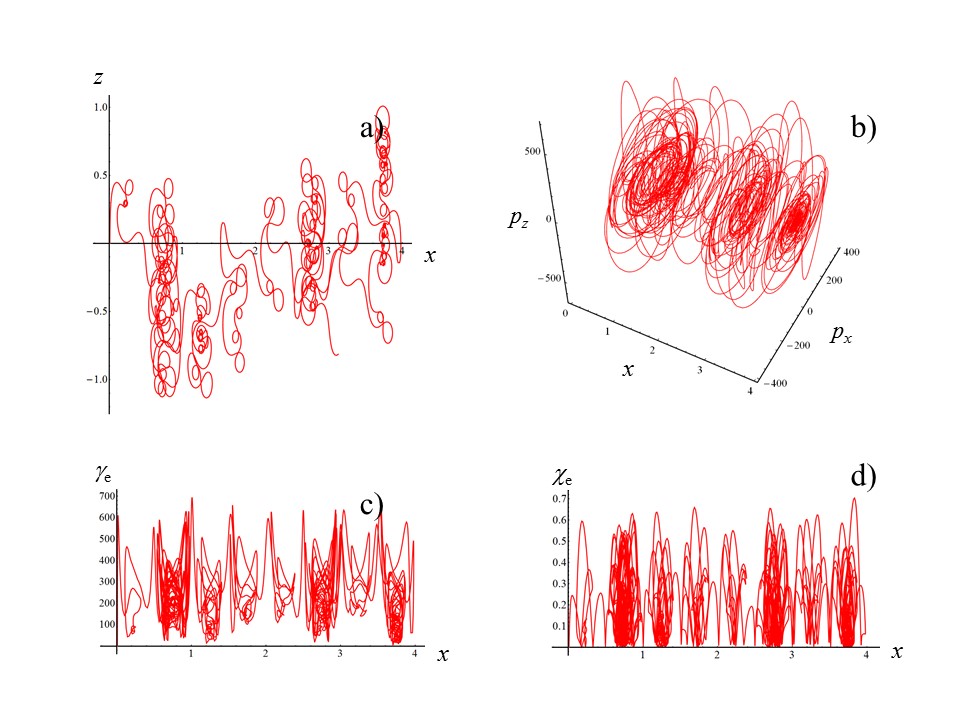}
		       \end{center}
 \caption{ a) Electron trajectories in the $(x,z)$ plane for initial conditions: $x(0)=0.01,\, z(0)=0,
 \, p_x(0)=0,\,p_z(0)=0$.
b) Trajectory in the phase space $x,p_x,p_z$;
c) Electron gamma-factor $\gamma_e$ versus the coordinate $x$;
d) Parameter $\chi_e$ versus the coordinate $x$, for the same initial conditions. 
The electromagnetic field amplitude is  $a_0=617$ 
 and the dissipation parameter is $\varepsilon_{rad}=1.2\times10^{-8}$. The coordinates, time and momentum
 are measured in the $2\pi c/\omega$, $2\pi/\omega$ and $m_e c$ units.
  \label{FIG-2}}
 \end{figure*}

	 In the limit of relatively weak dissipation, which corresponds to the domain I in Fig. \ref{Fig:a-chi_e}, the electron trajectory wanders in the phase space and in the 
	 coordinate space as shown in Fig. \ref{FIG-2}. 
	 In this case the wave amplitude is $a_0=618$.
	 The dissipation parameter equals $\varepsilon_{rad}=2\times10^{-8}$. 
	 The normalized critical QED field is $a_S=eE_S/m_e \omega c=m_ec^2/\hbar \omega=4\times 10^5$.
	 The parameter values correspond to the vicinity of the point $(a/a^*=1, \omega/\omega^*=1)$ in Fig. \ref{Fig:a-chi_e} b).
	 The integration time equals $75$.
	 
\begin{figure}
	   \begin{center}
    \includegraphics[width=6 cm]{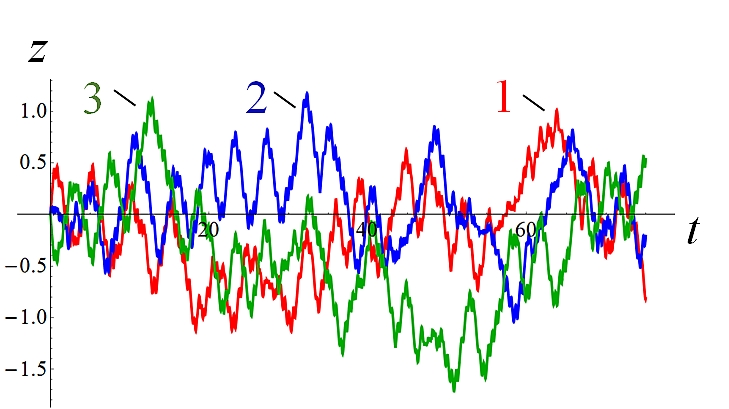}
		       \end{center}
 \caption{ Electron coordinate $z$  versus time space for initial coordinates $x(0)=0.01$--1,  $0.2$--2, $0.49$--3, other 
 parameters are the same as in Fig. \ref{FIG-2}.
  \label{FIG-3}}
 \end{figure}

	  Fig.  \ref{FIG-2} demonstrates a typical behavior of the electron in the limit of relatively 
	  low EM wave amplitude. Fig.  \ref{FIG-2} a) and b) show that the electron performs 
	 a random-walk-like motion for a long time being intermittently trapped and untrapped in the vicinities of the 
	 zero-electric field nodes, where the electric field vanishes. For this parameter choice 
	 the equilibrium trajectory at the electric field antinodes is unstable according to Ref. \cite{SSB-2010b} (see also \cite{Gong-Hu}).
	 The maximum value of the electron gamma-factor, $\gamma_e$, whose 
	 dependence on the coordinate $x$ is plotted in Fig.  \ref{FIG-2} c),  reaches 700. 
	 In the oscillating electric field of amplitude $a=618$ it would be equal to 618. The parameter $\chi_e$ (see Fig.  \ref{FIG-2} d))
	  changes between zero and approximately 0.7, which corresponds within an order of magnitude to 
	  $(a_0/a_S)\gamma_e$.   The particle coordinates $z$  versus time space in Fig. \ref{FIG-3} for initial coordinates  $x(0)=0.01$--1,  $0.2$--2, $0.49$--3 with other  
	  parameters the same as in Fig. \ref{FIG-2} show their wandering along the coordinate $z$. 
	  The particle over-leaping from one field period to another with small scale oscillations in between seen in Fig. \ref{FIG-3} may correspond to 
 L\'evy flights (see \cite{LEVYFLIGHTS, LEVYFLIGHTS1, LEVYFLIGHTS2, LEVYFLIGHTS3}).

 
\begin{figure}
	   \begin{center}
   \includegraphics[width=6 cm]{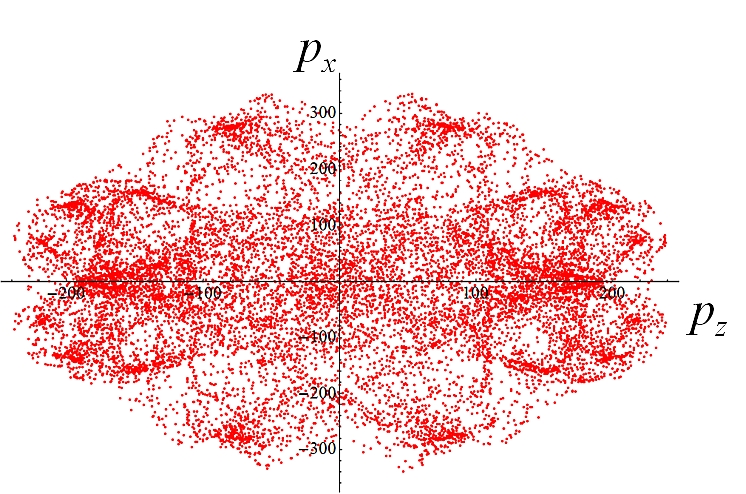}
		       \end{center}
 \caption{ The Poincar\'e sections showing
the particle positions in the phase plane ($p_x,\,p_z$) at  discrete times with
the time step equal to the period of the driving force.  The parameters are the same 
as in Fig. \ref{FIG-2} for $x(0)=0.01$.
  \label{FIG-4}
  }
 \end{figure}

Fig. \ref{FIG-4}, shows the Poincar\'e section for the motion of 
the particle with $x(0)=0.01$ positions in the phase plane $(p_x,p_z$) at 
discrete times with the time step equal to the period of the
driving force.  The parameters are the same 
as in Fig. \ref{FIG-2}. The Poincar\'e section demonstrates that this process is stochastic. 

   \begin{figure}
	   \begin{center}
    \includegraphics[width=8 cm]{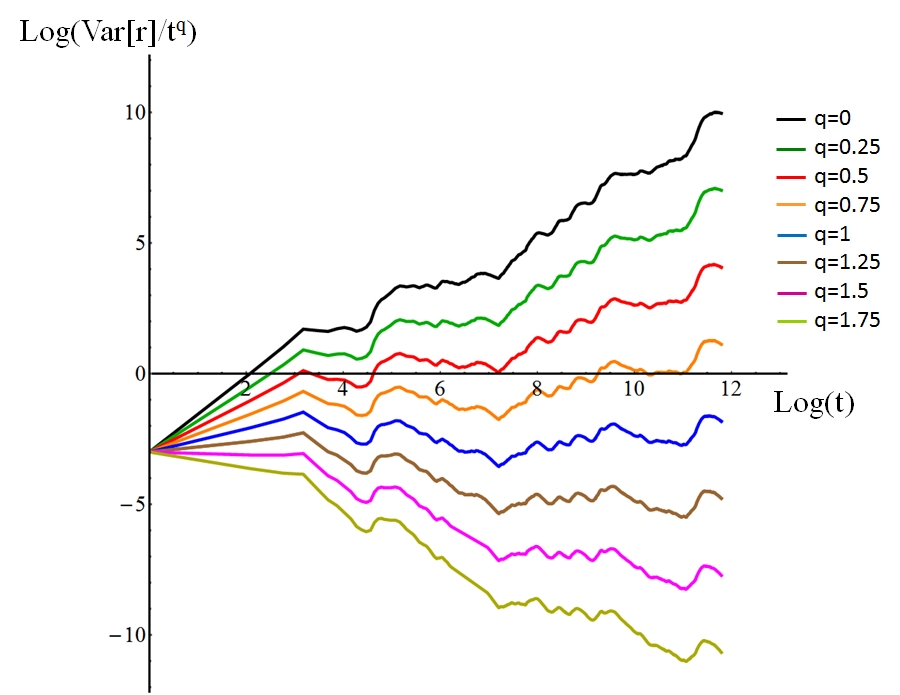}
		       \end{center}
 \caption{ Dependences of ${\rm Log}({\rm Var}[r]/t^q)$ on ${\rm Log}(t)$ for $0<q<1.25$ for the parameters corresponding to Fig. \ref{FIG-2}.
  \label{FIG-3bis}}
 \end{figure}
 
 \subsection{Random walk}
  
 Now we analyze the time dependence of the random walk, assuming that the particle coordinates $x(t)$  and $z(t)$  
  are  random variables, i.e. the particle displacement in the $(x,z)$ plane equal to $r=\sqrt{x^2+z^2}$ is also a random variable.  
  As is known in statistics the  behavior of the random variable $f$ is characterized by 
  the expectation $\mu=E[f]$ and variance $\sigma^2={\rm Var}[f]$ defined as 
  \begin{equation}
  E[f]=\lim_{t\to \infty}\frac{1}{t}\int^t f(t)dt
  \label{eq:expect}
    \end{equation}
  and
  \begin{equation}
{\rm Var}[f]=E[(f-E[f])^2].
  \label{eq:variance}
    \end{equation}
  The definition of an expectation in the form (\ref{eq:expect}) implies that the probability density function is taken to
  be a continuous uniform distribution equal  to $1/t$ within the interval $[0,t]$.  We assume here that the ergodicity of the processes is expected.
  
  If the random walk process is a Wiener process, which also called ``Brownian motion'', the  variance of the walker's coordinate $r(t)$ is proportional to time (e.g. see \cite{Durrett}).
To  examine whether or not the random walk seen in Figs. \ref{FIG-2} and \ref{FIG-3} is a Wiener process we plot in Fig. \ref{FIG-3bis} the dependences of ${\rm Log}({\rm Var}[r]/t^q)$ on ${\rm Log}(t)$ for $0<q<1.75$. For the Wiener process the parameter $q$ should be equal to $1$. As we can see, in our case random walk process the variance is proportional 
to $t^q$ with $q\approx 1$.

  \begin{figure*}
	   \begin{center}
   \includegraphics[width=12 cm]{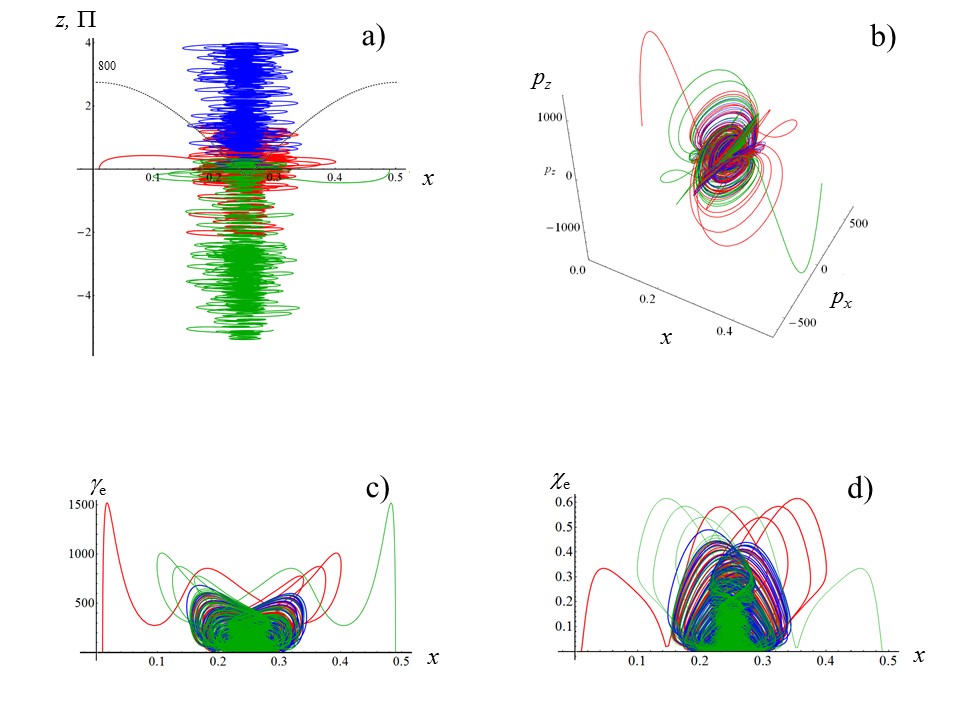}
		       \end{center}
 \caption{ Electron motion in the standing EM wave for $\varepsilon_{rad}=6\times 10^{-9}$, $a_S=8\times 10^5$,  and  $a_0=778$ 
 for initial conditions: $x(0)=0.01,\, z(0)=0, \, p_x(0)=0,\,p_z(0)=0$ (red); 
$x(0)=0.2,\, z(0)=0, \, p_x(0)=0,\,p_z(0)=0$ (blue); $x(0)=0.49,\, z(0)=0, \, p_x(0)=0,\,p_z(0)=0$ (green).
a) Trajectory in the $x,z$ plane. 
Dashed line is the ponderomotive potential (\ref{eq:pndpot}) vs the $x$ coordinate;
b) Electron trajectories in the $(x,p_x, p_z)$ space.  
c) Electron gamma-factor $\gamma_e$ versus the coordinate $x$.
d) Parameter $\chi_e$ versus the coordinate $x$, for the same initial conditions.
  \label{FIG-5}
  }
 \end{figure*}

 \subsection{Moderate intensity regime}
 
 The situation qualitatively changes, when the dissipation becomes more significant. In Fig. \ref{Fig:a-chi_e} 
 this corresponds to the domain II. 
  This case is illustrated in Fig. \ref{FIG-5},
 for which the radiation friction parameter is $\varepsilon_{rad}=6\times 10^{-9}$, 
 the normalized critical QED field is $a_S=8\times 10^5$, and 
the normalized laser field equals  $a_0=778$. In Fig. \ref{FIG-5} we present three trajectories  
 for particles with initial conditions: $x(0)=0.01,\, z(0)=0, \, p_x(0)=0,\,p_z(0)=0$ (red); 
$x(0)=0.2,\, z(0)=0, \, p_x(0)=0,\,p_z(0)=0$ (blue); $x(0)=0.49,\, z(0)=0, \, p_x(0)=0,\,p_z(0)=0$ (green).
As seen in Fig. \ref{FIG-5} a), where the trajectories in the $x,z$ plane are shown, 
independent of the initial position all three trajectories 
end up in the vicinity of the plane $x=0.25$. Here the EM wave electric field vanishes. 

\begin{figure}
	   \begin{center}
   \includegraphics[width=6 cm]{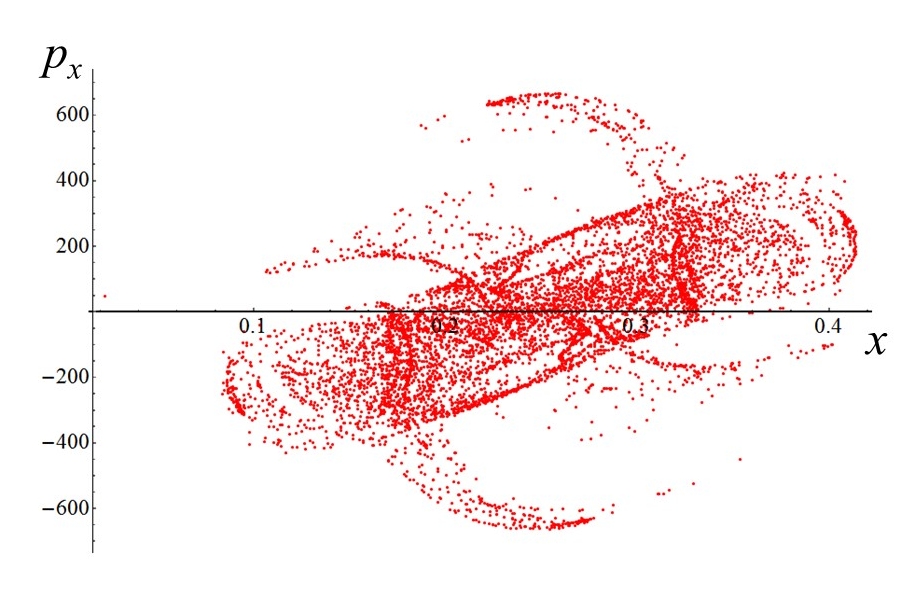}
		       \end{center}
 \caption{ The Poincar\'e sections showing
the particle positions in the phase plane ($x,p_x$) at discrete times with
the time step equal to the period of the driving force. The parameters are the same 
as in Fig. \ref{FIG-5} for $x(0)=0.01$.
  \label{FIG-6}
  }
 \end{figure}

At the coordinate $x=0.25$ the ponderomotive potential has a minimum. It is  defined as 
	  \begin{equation}
 \label{eq:pndpot}
 \Pi(x)=\frac{1}{2 \pi}\int_{-\pi}^{\pi}(\sqrt{1+A(x,t)^2}-1)dt
 \end{equation}
  with the vector potential $A(x,t)$ given by Eq. (\ref{eq:4-poten2waves}).
The dashed curve in Fig. \ref{FIG-5} a) presents the ponderomotive potential (\ref{eq:pndpot}) 
dependence on the $x$ coordinate.
In Fig. \ref{FIG-5} b) electron trajectories in the $(x,p_x, p_z)$ 
space show the attractors, which  have been analyzed in details in Ref. \cite{Esirkepov-2015} 
(see Fig. \ref{FIG-6}, where the Poincar\'e section is presented for this case).   
Electron gamma-factors $\gamma_e$ versus the coordinate $x$ presented in Fig. \ref{FIG-5} c) 
correspond to the case when the dissipation limits the particle energy, which does not exceed 
the value determined by the amplitude of the EM wave being of the order of $a$. 
Since the parameter $\chi_e(x)$  plotted in Fig. \ref{FIG-5} d) is less than unity for 
all three trajectories, the equation of an electron motion with the radiation friction force is still valid for this parameter range.
 
 In Fig. \ref{FIG-6}, we plot the Poincar\'e section for 
the particle with the same  parameters
as in Fig. \ref{FIG-5} for $x(0)=0.01$. Here are the particle positions in the phase plane ($x,p_x$) at
discrete times with the time step equal to the period of the
driving force. The map pattern corresponds to the stochastic regime developed in 
the particle motion.

  \begin{figure*}
	   \begin{center}
   \includegraphics[width=12 cm]{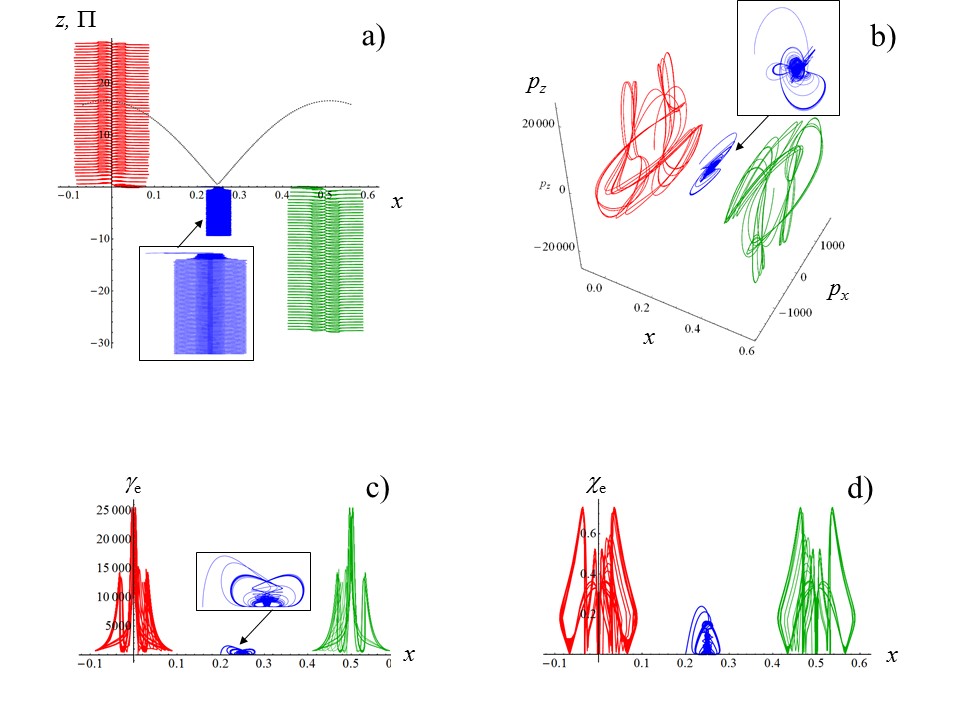}
		       \end{center}
 \caption{ Electron motion in the standing EM wave for $\varepsilon_{rad}=1.2\times 10^{-9}$, $a_S=4\times 10^6$, $a=1996$ 
 for initial conditions: $x(0)=0.01,\, z(0)=0, \, p_x(0)=0,\,p_z(0)=0$ (red); 
$x(0)=0.2,\, z(0)=0, \, p_x(0)=0,\,p_z(0)=0$ (blue); $x(0)=0.49,\, z(0)=0, \, p_x(0)=0,\,p_z(0)=0$ (green).
a) Trajectory in the $x,z$ plane. Inset shows zoomed trajectory for $x(0)=0.2$. Dashed line is the ponderomotive potential (\ref{eq:pndpot}) vs the $x$ coordinate;
b) Electron trajectories in the $(x,p_x, p_z)$ space.  Inset shows zoomed trajectory for $x(0)=0.2$ corresponding to a strange attractor \cite{Esirkepov-2015}.
c) Electron gamma-factor $\gamma_e$ versus the coordinate $x$. Inset shows zoomed $\gamma_e(x)$ for $x(0)=0.2$
d) Parameter $\chi_e$ versus the coordinate $x$, for the same initial conditions.
  \label{FIG-7}
  }
 \end{figure*}

 \subsection{High intensity regime}
 	
 If we choose the parameters in a such the way that the dissipation becomes even more significant, when we approach 
 the domain IV in Fig. \ref{Fig:a-chi_e}, the particle 
 behavior becomes counterintuitive, as can be seen  in Fig.~\ref{FIG-7},
 for which the radiation friction parameter is $\varepsilon_{rad}=1.2\times 10^{-9}$, 
 the normalized critical QED field is $a_S=4\times 10^6$, and 
the normalized laser field equals  $a_0=1996$. There we present three electron trajectories  
 for the same initial conditions as in Fig. \ref{FIG-4}: $x(0)=0.01,\, z(0)=0, \, p_x(0)=0,\,p_z(0)=0$ (red); 
$x(0)=0.2,\, z(0)=0, \, p_x(0)=0,\,p_z(0)=0$ (blue); $x(0)=0.49,\, z(0)=0, \, p_x(0)=0,\,p_z(0)=0$ (green).
Trajectories in the $x,z$ plane (Fig. \ref{FIG-5} a)) are principally different depending on where 
the particle has been initially localized. For $x(0)=0.2$ the trajectory remains in the vicinity of the ponderomotive 
potential minimum similarly to the case discussed above (Fig. \ref{FIG-5}). 
The dashed line is the ponderomotive potential (\ref{eq:pndpot}) vs the $x$ coordinate.
In contrast, particles with initial coordinates near the maximum of the ponderomotive potential are trapped 
there (similar behavior was noted in Ref. \cite{Gonoskov-2014}). 

In Fig. \ref{FIG-7} 
b) the electron trajectories in the $(x,p_x, p_z)$ space show behaviour typical for limit circles and attractors. 
The inset shows the zoomed trajectory for $x(0)=0.2$ corresponding to a strange attractor \cite{Esirkepov-2015}. 
The trajectories with $x(0)=0.01$ and $x(0)=0.49$ demonstrate regular limit circles.
It follows from Fig. \ref{FIG-7} c) that the electron gamma-factor 
$\gamma_e$ has a moderate value for the electron trapped near the ponderomotive potential minimum 
(the inset shows zoomed $\gamma_e(x)$ for $x(0)=0.2$), 
and the particles are efficiently accelerated when they are trapped in the region 
at the ponderomotive potential maximum.
For the parameters chosen  $\chi_e(x)$  plotted in Fig. \ref{FIG-7} d) remains less than unity for all three trajectories, 
i. e. the QED effects are finite but relatively weak.

\begin{figure}
	   \begin{center}
   \includegraphics[width=6 cm]{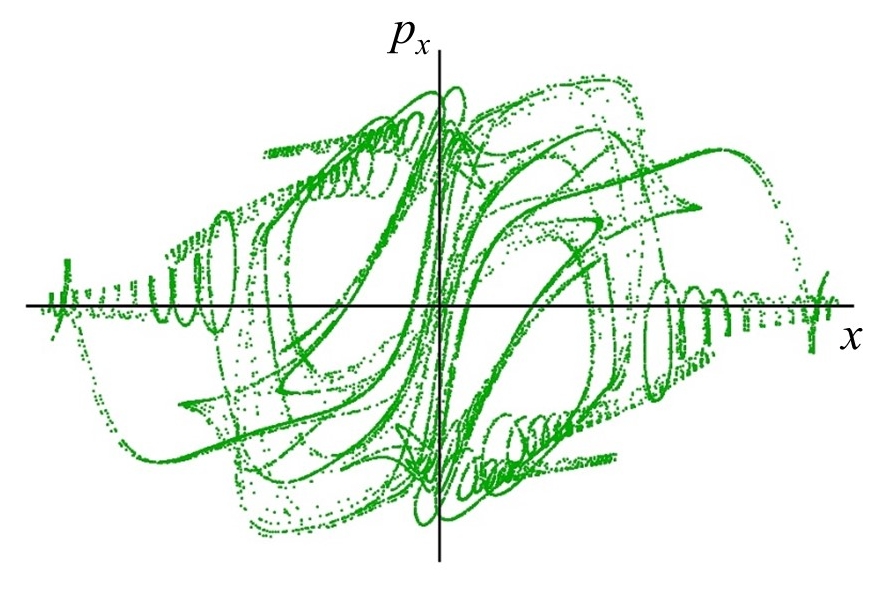}
		       \end{center}
 \caption{ The Poincar\'e sections showing
the particle positions in the phase plane ($x,p_x$) at discrete times with
the time step equal to the period of the driving force. The parameters are the same 
as in Fig. \ref{FIG-7} for $x(0)=0.49$.
  \label{FIG-8}
  }
 \end{figure}

In Fig. \ref{FIG-8}, we plot the Poincar\'e section showing
the particle with $x(0)=0.49$ positions in the phase plane ($x,p_x$) at
discrete times with the time step equal to the period of the
driving force.  The parameters are the same 
as in Fig. \ref{FIG-7} for $x(0)=0.49$. Although the map pattern is pretty complicated it does 
not contain curve broadening, i.e. does not indicate a stochastic regime 
of the particle motion.

In the next Section we discuss the mechanism of dissipative particle trapping  
in the vicinity of the ponderomotive potential maximum, which can explain the observed effects (see also \cite{FEDOTOV-2014}).

\section{Simple model of the stabilization of the particle motion in an oscillating 
field due to  the nonlinear friction}
 
Let us consider a particle motion in a fast oscillating field in a way similar to \cite{LLM}. 
As in Ref. \cite{LLM} for the sake of simplicity of calculations we assume non-relativistic electron motion in one dimension,  
when the force acting on the particle depends on the coordinate $x$ and time $t$. In contrast to the consideration 
in Ref. \cite{LLM}, we take into account the effects of the friction. The equation of particle motion is
	  \begin{equation}
 \label{eq:equmot1D}
\ddot x+\kappa(F) \dot x=F.
 \end{equation}
Here a dot stands for the time derivative and $\kappa(F)$ is the  friction coefficient. It is assumed  to depend on the rapidly oscillating driving force, 
	  \begin{equation}
 \label{eq:driverF}
F(x,t)=f_1(x)\cos \omega t+f_2(x)\sin \omega t.
 \end{equation}

We seek a solution of Eq. (\ref{eq:equmot1D}), assuming that it 
can be written down as a sum of two parts, 
	  \begin{equation}
 \label{eq:X-xi}
x(t)=X(t)+\xi(t) \, ,
 \end{equation}
where $X(t)$ slowly varies with time and
$\xi(t)$ is a
small fast oscillating periodic function, $|\xi|\ll |X|$.

We also assume that the time average of the function $\xi(t)$ 
over the oscillation period $2 \pi/\omega$ is zero.
Introducing the notation
\begin{equation}
 \label{eq:ave-brackets}
\left< x \right>=\frac{\omega}{2\pi}\int_{0}^{{2\pi}/{\omega}} x(t) dt \, ,
 \end{equation}
we obtain
\begin{equation}
 \label{eq:ave-xi}
\left< \xi \right>=\left<\dot\xi \right>=\left< \ddot\xi \right> =0 \, .
 \end{equation}
Therefore, we have $\left< x\right>=X(t)$, 
i.e. the function $X(t)$ describes the slow particle motion averaged over the fast oscillations,
 $\left< X\right>\approx X(t)$.

Substituting (\ref{eq:X-xi}) into the equation of particle motion (\ref{eq:equmot1D}) and 
expanding the functions $\kappa(x,t)$ and $F(x,t)$ in powers of $\xi$, i. e. writing 
$\kappa(x,t)\approx \kappa(X,t)+\xi \partial_X \kappa(X,t)$
and
$F(x,t)\approx F(X,t)+\xi \partial_X F(X,t)$,
we obtain
	  \begin{equation}
 \label{eq:equmot1DXxi}
 \ddot X+ \ddot \xi+\kappa \dot X+\kappa \dot \xi+
\xi \dot X \partial_X \kappa + \xi \dot \xi \partial_X \kappa
=F+\xi \partial_X F,
 \end{equation}
where $\partial_X$ is the partial derivative with respect to the first argument of
functions $\kappa(X,t)$ and $F(X,t)$.
This equation contains slowly varying and fast oscillating terms, which apparently should be separately equal to each other.
In the zeroth order approximation 
with respect to small function $\xi$ and the time derivatives of the slow function $X$,
we find the equation for the fast oscillating term
\begin{equation}
 \label{eq:equmot-osc}
\ddot \xi+\kappa \dot \xi=F \, .
\end{equation}
Here we neglect the terms proportional to $\xi$.
The time derivatives $\ddot\xi$ and $\dot\xi$
are not small, being proportional to $\omega^2$ and $\omega$, respectively.
They are assumed to be much greater than $\ddot X$ and $\dot X$.
The friction coefficient $\kappa$ is not necessarily small.

Integration of Eq. (\ref{eq:equmot-osc}), assuming $X$ to be constant, yields
\begin{eqnarray}
\xi &=& \xi_0 + \int_{0}^{t}d\tau
\left[
e^{-K(X,\tau)}\dot\xi_0 + 
\right.
\nonumber
\\
&&\left.
\int_{0}^{\tau} e^{K(X,\tau')-K(X,\tau)} F(X,\tau') d\tau'
\right]  \, ,
\\
&&K(X,t)=\int_0^t\kappa(X,\tau)d\tau \, .
\label{eq:equmot-osc-int}
\end{eqnarray}
Assuming that $K(X,t)$ can be approximated by
$K(X,t) \approx \left<\kappa\right> t $,
where $\left<\kappa\right>$ is the time-averaged friction coefficient,
in the limit $t\gg 1/\left<\kappa\right>$ we obtain
\begin{eqnarray}
\xi&=&\frac{(\left<\kappa\right> f_1-\omega f_2) \sin\omega t
-(\left<\kappa\right> f_1+\omega f_2) \cos\omega t}{\omega(\left<\kappa\right>^2+\omega^2)}
\nonumber
\\
&&=-\frac{1}{\left<\kappa\right>^2+\omega^2}
\left(F+\frac{\left<\kappa\right>}{\omega^2} \partial_t F \right)
\label{eq:equmot-osc-solve}
 \end{eqnarray}
with $ \partial_t F=\left.  \partial F(X,t)/\partial t\right|_{X={\rm const}} $.
Here we assumed the initial condition $\xi_0 = -(f_2+\omega\dot\xi_0)/(\left<\kappa\right>\omega)$.

Averaging Eq. (\ref{eq:equmot1DXxi}) over time
and  taking into account that $\left< F(X,t)\right>\approx 0$ for nearly constant $X(t)$, we obtain
\begin{equation}
\label{eq:equmot1DX}
\ddot X+(\left<\kappa\right>+\left<\xi\partial_X\kappa\right>) \dot X=
\left< \xi \partial_X F \right> -
\left< \xi \dot \xi \partial_X \kappa\right>
\, .
\end{equation}

Substituting expression (\ref{eq:equmot-osc-solve}) into the r.h.s. of Eq. (\ref{eq:equmot1DX}),
 for the first term we obtain
\begin{equation}
\label{eq:force1pond}
\left< \xi \partial_X F \right>=
-\frac{ \partial _X (f_1^2+f_2^2)}{4 (\left<\kappa\right>^2+\omega^2)}
-\frac{\left<\kappa\right>(f_2 \partial _X f_1 - f_1 \partial _X f_2)}{2 \omega(\left<\kappa\right>^2+\omega^2)} 
\, .
\end{equation}

The first term on the r.h.s. of Eq. (\ref{eq:force1pond}) is the well known ponderomotive force  \cite{LLM} where
the friction effect is taken into account.
The last term on the r.h.s., proportional to the friction coefficient, can 
change signs for $f_2 \partial _X f_1\ne f_1 \partial _X f_2$ 
depending on whether $f_2 \partial _X f_1> f_1 \partial _X f_2$ or $f_2 \partial _X f_1< f_1 \partial _X f_2$. 
It vanishes if $f_2 \partial _X f_1= f_1 \partial _X f_2$, 
$f_1\ne 0,\, f_2=0$ or $f_2\ne 0,\, f_1=0$.

The actual form of the last term on the r.h.s. of Eq. (\ref{eq:equmot1DX})
is determined by the specific dependence of the friction coefficient $\kappa$ on the driver force.
As an example, we consider the case when this dependence 
is quadratic, i.e. $\kappa=\nu F^2$, with a constant $\nu$. 
Then we obtain
\begin{eqnarray}
&&- \left<\xi \dot \xi \partial_X \kappa\right>=
\frac{\nu \left<\kappa\right> \partial_X(f_1^2+f_2^2)^2}{8 (\left<\kappa\right>^2+\omega^2)^2} +
\nonumber
\\
&&
\frac{\nu (\left<\kappa\right>^2-\omega^2)(f_1^2+f_2^2)(f_2 \partial _X f_1 - f_1 \partial _X f_2)}{4\omega (\left<\kappa\right>^2+\omega^2)^2}
 \label{eq:force2drag}
\end{eqnarray}
and the time-averaged friction coefficient becomes $\left<\kappa\right>=\nu(f_1^2+f_2^2)/2$.
In addition, $\left<\xi\partial_X\kappa\right>=0$.

For the sake of simplicity we further assume that $f_2=0$ 
in expression (\ref{eq:driverF}) for the driver force.  
Then, Eqs. (\ref{eq:force1pond}) and (\ref{eq:force2drag}) are simplified and we finally obtain
the equation for the slowly varying function $X(t)$:
\begin{equation}
\label{eq:equmot1DX-bis}
\ddot X+\frac{\nu f_1^2}{2} \dot X=
-\frac{ \partial _X f_1^2}{\nu^2 f_1^4+4\omega^2}
+\frac{2\nu^2 \partial_X f_1^6}{3(\nu^2 f_1^4+4\omega^2)^2}
\, .
\end{equation}
The first term on the r.h.s. corresponds to the ponderomotive force, the last term
is the drag force induced by the friction.

As we can see, the ponderomotive force (\ref{eq:force1pond}) and the drag force due to the friction  (\ref{eq:force2drag})  have different signs in Eq. (\ref{eq:equmot1DX-bis}).
If 
\begin{equation}
|\nu|>2\omega/f_1^2, 
\label{eq:criterion}
\end{equation}
the drag force exceeds in magnitude the ponderomotive force. 
Using Eq. (\ref{eq:fradQ}) for radiation friction force and the equation (\ref{eq:equmot-mom}) 
of electron motion we can find that the criterion (\ref{eq:criterion}) can be written as the condition 
on the laser amplitude, $a\,\varepsilon^{1/3}_{rad}>1$., i.e. the drag force due to the radiation friction 
is larger than the ponderomotive force in the radiation friction dominated regime (see Ref. \cite{FEDOTOV-2014}).

\begin{figure*}
	   \begin{center}
   \includegraphics[width=10 cm]{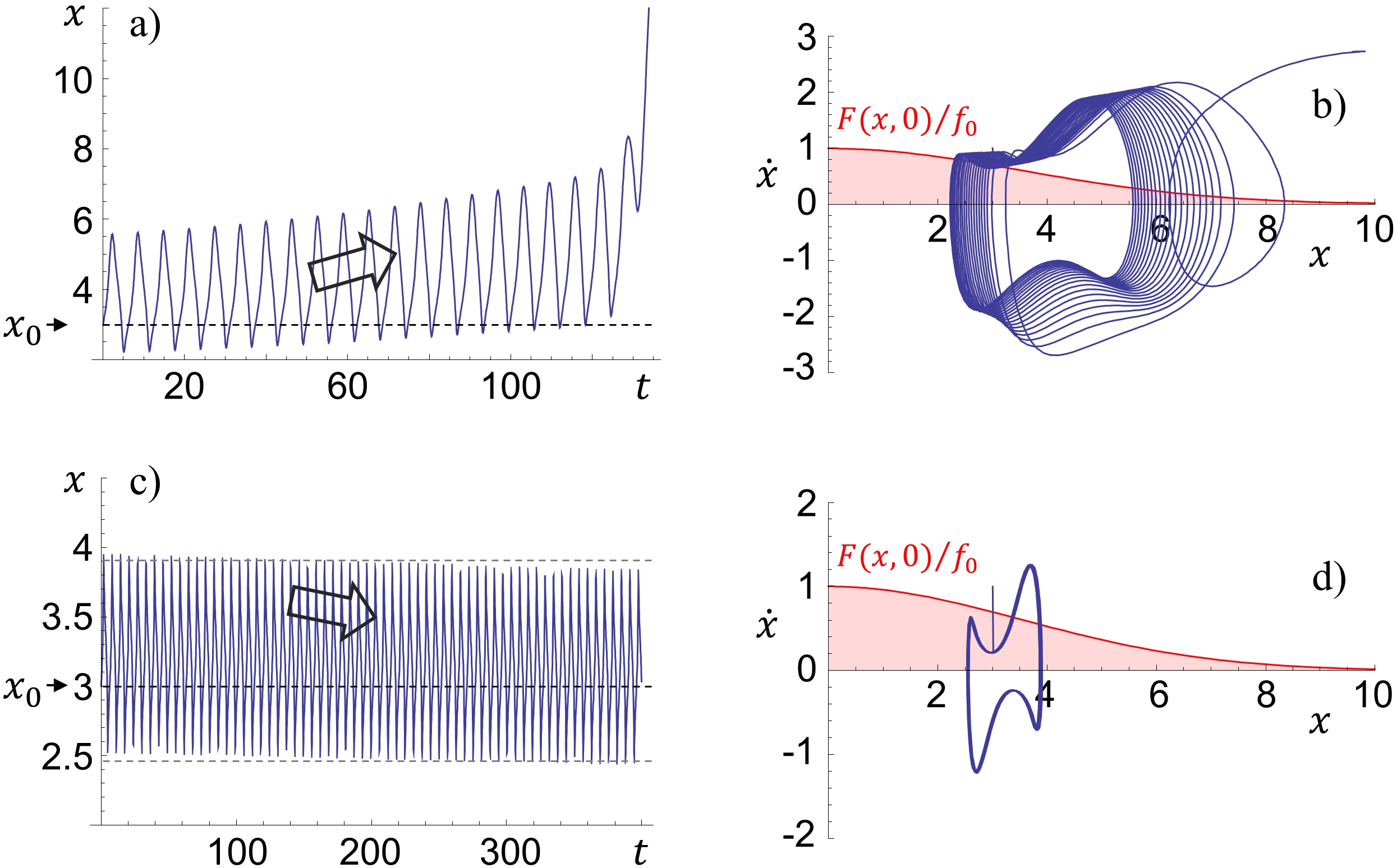}
		       \end{center}
 \caption{ The solutions of Eq. (\ref{eq:equmot1D}) in the case of relatively weak driver force ( a and b), and 
 for the case of relatively strong driver force (c and d). a) and c) Dependence of the particle coordinate on time. b) and d) The  particle trajectory in the phase plane $(x,\dot x)$.
  \label{FIG-9}
  }
 \end{figure*}

Numerical integration of the equation of motion (\ref{eq:equmot1D}) with 
\begin{equation}
F(x,t)=f_0\exp(-(x/l_0)^2) \cos \omega t \quad {\rm and} \quad \kappa=\nu F^2
\end{equation}
 reveals the main features of the behavior predicted 
within the framework of the simple model approximation. 
The solutions for the cases of relatively weak and relatively strong driver force are plotted in Fig.\,\ref{FIG-9}. The 
parameters are as follows. The driver frequency and the friction coefficient values are $\omega=1$ and $\nu=0.1$,
respectively. The driver width equals $l_0=5$. The initial coordinate and velocity are $x_0=3$ and $\dot x_0=1$, 
in both cases. 
The driver amplitude is equal to $f_0=5\sqrt{2 \omega/ \nu}$ in the case of the weak driver, 
and is equal to $f_0=15\sqrt{2 \omega/ \nu}$ in the case 
of the strong driver. As we see in Fig.\,\ref{FIG-9} a) and b), in the case of weak nonlinearity, 
the particle being pushed outwards by the ponderomotive force having performed several 
oscillations leaves the region where the driver 
force is localized. In contrast, in the limit of strong nonlinearity, the friction drag force prevents the particle from 
leaving the driver localization region resulting in its slow drift inwards (Fig.\,\ref{FIG-9} c) and d)).
 
On a trajectory corresponding to a quasi-periodic particle motion seen in Fig.\,\ref{FIG-9},
the particle feels an almost constant driving force.
This situation can be described in the approximation
\begin{equation}
F(x,t)=f_0 \cos \omega t, \qquad 
\kappa=\nu f_0^2 \cos^2 \omega t. 
\end{equation}
Substituting this driving force and friction coefficient into Eq. (\ref{eq:equmot1D}),
we change variables to $(\tau,y(\tau))$,
$t = \tau/\omega$, $x(t) = (f_0/\omega^2)y(\tau)$
and introduce the constant 
\begin{equation}
\sigma=\nu f_0^2/\omega.
\end{equation}
Thus, we obtain
\begin{equation}
y''(\tau) + \sigma \cos^2(\tau) y'(\tau) = \cos \tau
\, .
 \label{eq:equmot1D:fconst}
\end{equation}
Here a prime denotes a differentiation with respect to the variable $\tau$. Using  Eq. (\ref{eq:equmot-osc-int})
and the generating function for the modified Bessel functions of the first kind,
$I_k$, one can cast the general solution to Eq. (\ref{eq:equmot1D:fconst})
in the form
\begin{equation}
y'(\tau)=
\exp\left({-\frac{\sigma \tau}{2}-\frac{\sigma}{4}\sin(2\tau)}\right)
[ y'(0) - Y_{LC}(0) ] +  Y_{LC}(\tau)
\, ,
 \label{eq:equmot1D:fconst:sol}
\end{equation}
where the function $Y_{LC}(\tau)$ is given by 
\begin{eqnarray}
Y_{LC} (\tau) &=&
\textstyle
\exp\left({-\frac{\sigma}{4}\sin(2\tau)}\right)
\sum\limits_{k=-\infty}^{\infty}
   (-i)^{k+1}  I_k\left(\frac{\sigma}{4}\right)
\nonumber \\
\textstyle
&&   \times \left\{
    \frac{\exp\left[{i(2k-1)\tau}\right]}{4k-2-i\sigma}
 + \frac{\exp\left[{i(2k+1)\tau}\right]}{4k+2-i\sigma}
   \right\}. \,\,\,\,\,\quad
\label{eq:equmot1D:fconst:LC}
\end{eqnarray}
As one can see, any solution at $\tau\rightarrow\infty$ 
tends to the limit cycle described by the function $Y_{LC}$
and determined by the constant $y'(0)=Y_{LC} (0) $.

\begin{figure}
\begin{center}
   \includegraphics[width=8 cm]{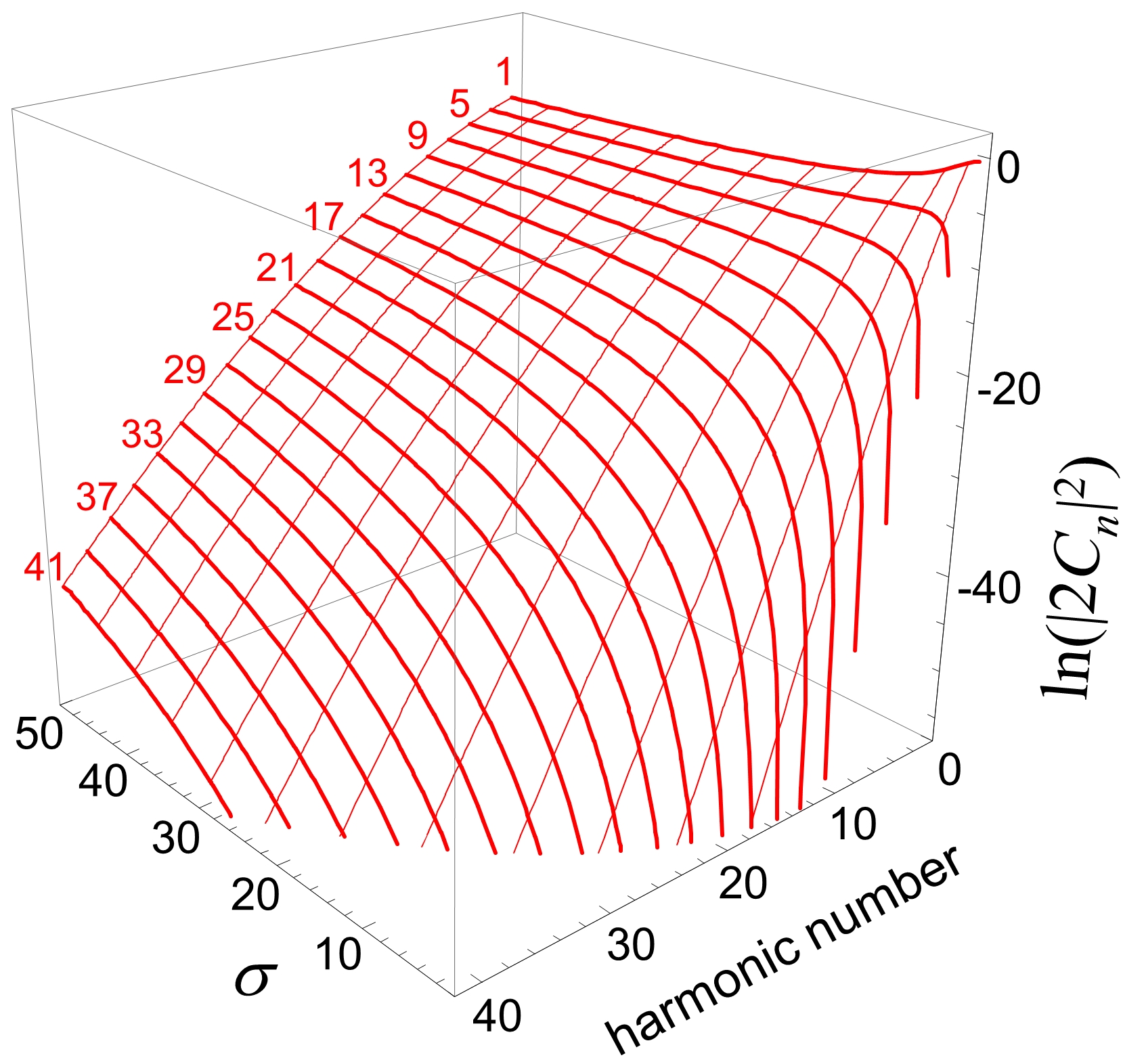}
\end{center}
 \caption{ The spectral density of the particle velocity 
for several harmonics of the driving force frequency
as a function of the friction parameter $\sigma$.
\label{FIG-9bis}
}
 \end{figure}

The function describing the limit cycle, Eq. (\ref{eq:equmot1D:fconst:LC}),
can be represented as a Fourier series
in terms of odd harmonics of the driving force frequency
\begin{eqnarray}
Y_{LC}(\tau)=&&\sum\limits_{n=1}^{\infty}
\left[
\exp\left({i(2n-1)\tau}\right)C_{n}(\sigma)\right .
\, \nonumber \\
&&
\left .+\exp\left({-i(2n-1)\tau}\right)C_{1-n}(\sigma)
\right]
\end{eqnarray}
with
\begin{equation}
C_n(\sigma) =
\textstyle
i^n
\sum\limits_{k=-\infty}^{\infty}
\frac{(-1)^{k+1}}{4k+2-i\sigma}
\left[
I_k\left(\frac{\sigma}{4}\right) - i I_{k+1}\left(\frac{\sigma}{4}\right)\right]
I_{k+1-n}\left(\frac{\sigma}{4}\right)
\, .
\label{eq:equmot1D:fconst:LC:F}
\end{equation}
This gives the spectrum of the limit cycle trajectory.
For the particle velocity (corresponding to $y'(\tau)$),
the spectral density is $|2C_{n}(\sigma)|^2$,
Fig. \ref{FIG-9bis}. 

\section{Regular and chaotic electron motion in three s- and p-polarized colliding laser pulses}

\subsection{EM field configuration}

Let us consider three s(p)-polarized waves, which $z$-component of the electric (magnetic) field is given by 
\[ \binom{{E_z}}{{B_z}}
=-\frac{1}{\sqrt{3}} \binom{{E_0}}{{B_0}}\left\{\sin\left [\omega_0 \left (t+\frac{ y}{c}\right)\right] \right .\]
\begin{equation}
\label{eq:EBzfield3waves}
\left .+2 \sin\left [\omega_0 \left (t-\frac{ y}{2 c}\right)\right]\cos\left (\omega_0\frac{ \sqrt{3} x}{2c}\right)\right\}.
\end{equation}
The $x$ and $y$ components of  the magnetic (electric) field of the s(p)-polarized wave  are
\[  \binom{{B_x}}{{E_x}}
=\frac{1}{\sqrt{3}} \binom{{\,\,\,E_0}}{{-B_0}} \left\{ -\sin\left [\omega_0 \left (t+\frac{ y}{c}\right)\right] \right .\]
\begin{equation}
\label{eq:EBfield3waves}
 \left .+\sin\left [\omega_0 \left (t-\frac{ y}{2 c}\right)\right]\cos\left (\omega_0\frac{ \sqrt{3} x}{2c}\right)\right\}
\end{equation}
and
\begin{equation}
\label{eq:BEfield3waves}
\binom{{B_y}}{{E_y}}= \frac{1}{\sqrt{3}} \binom{{\,\,\,E_0}}{{-B_0}}
\cos\left [\omega_0 \left (t-\frac{ y}{2 c}\right)\right]\sin\left (\omega_0\frac{ \sqrt{3} x}{2c}\right).
\end{equation}
\begin{figure}
\centering
\includegraphics[width=5 cm]{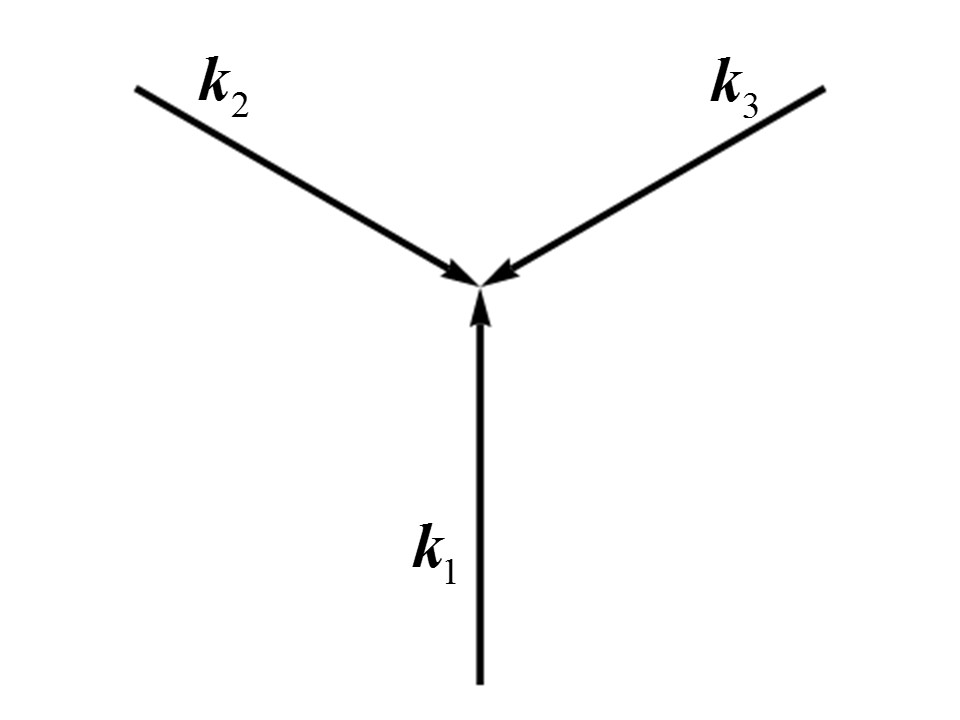}
\caption{ Wave vectors of three colliding waves.}
\label{FIG-10}
\end{figure}

\begin{figure*}
\centering
\includegraphics[width=8 cm]{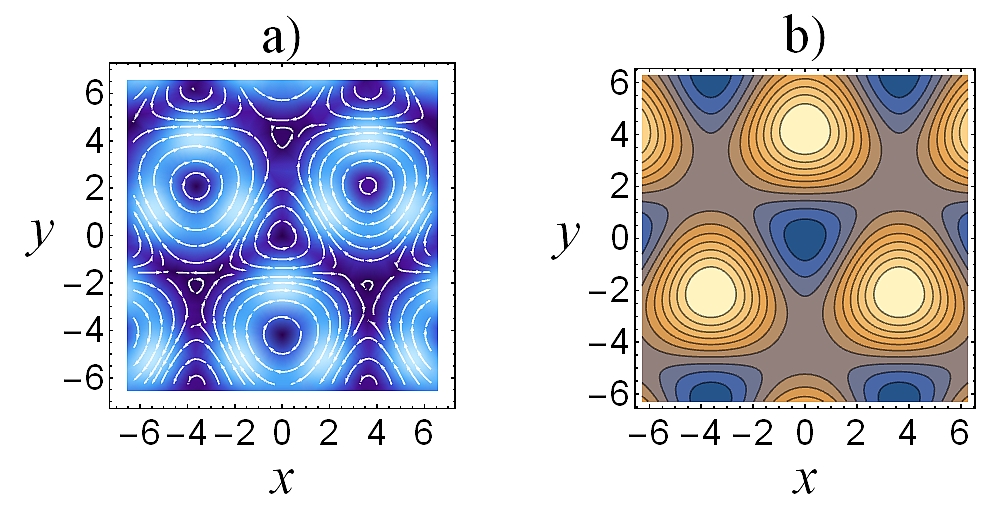}
\caption{ Three s-polarized (p-polarized) EM waves: a) magnetic (electric) field; 
b) isocontours of the electric (magnetic) field in the $(x,y)$ plane at time $t=\pi/4$.}
\label{FIG-11}
\end{figure*}

The wave orientation is illustrated in Fig.\,\ref{FIG-10}. 
As an example in Fig. \ref{FIG-11} a) we show the magnetic (electric) field  
${\bf B}_n=B_x{\bf e}_x+B_y{\bf e}_y\,\,\, ({\bf E}_n=E_x{\bf e}_x+E_y{\bf e}_y)$  and 
in Fig. \ref{FIG-11} b) the isocontours of the electric (magnetic) field $E_z$  ($B_z$) in the $(x,y)$ plane at time $t=\pi/4$ for 
the case of three colliding s-polarized (p-polarized) EM waves.

\subsection{Electron interaction with three s-polarized 
 EM waves}
 
   \subsubsection{Particular solutions}
 
 Due to the symmetry of the EM field given by Eqs. (\ref{eq:EBfield3waves},\, \ref{eq:BEfield3waves}), 
 there are particular solutions of the equations of motion, when the particle moves straight in the $(x,y)$ plane 
 along the direction of one of the waves propagation. If we let $x=0$ in Eqs. (\ref{eq:EBfield3waves},\, \ref{eq:BEfield3waves}), 
 the electromagnetic field formally corresponds to a superposition of two EM waves one of which 
 propagates with the velocity equal to $-c$ and another has the velocity $2 c$.

\begin{figure*}
	   \begin{center}
    \includegraphics[width=12 cm]{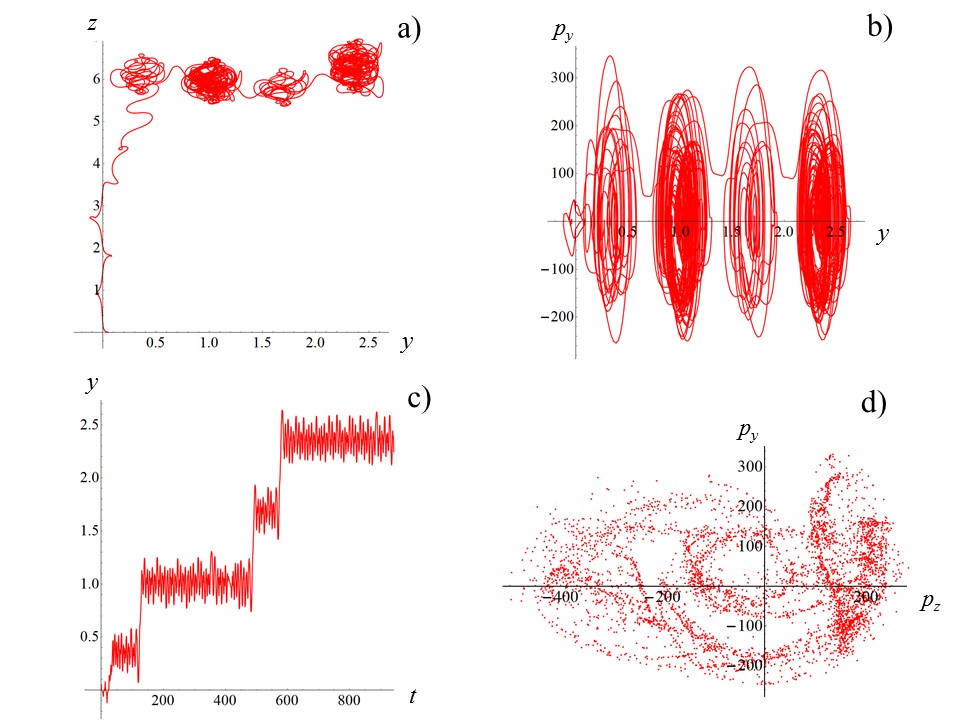}
		       \end{center}
 \caption{  a) Electron trajectory in the $(y,\,z)$ plane for initial conditions: $x(0)=0,y(0)=0.05,\, z(0)=0,
 \, p_x(0)=0,\,p_y(0)=0,\,p_z(0)=0$.
b) Trajectory in the phase $(y\,,p_z)$ plane;
c) Electron $y$-coordinate versus time;
d) The Poincar\'e sections showing
the particle positions in the phase plane $(p_z,p_y)$ at discrete times with
the time step equal to the period of the driving force. 
The electromagnetic field amplitude is  $a_0=436$, 
 the dissipation parameter is $\varepsilon_{rad}=1.2\times10^{-8}$, and the normalized critical QED field is $a_S=4\times 10^5$. The coordinates, time and momentum
 are measured in the $2\pi c/\omega$, $2\pi/\omega$ and $m_e c$ units.
  \label{FIG-12}}
 \end{figure*}

 The integration of the equations of electron motion yields the particle trajectories presented in Fig.\,\ref{FIG-12}
 for initial conditions: $x(0)=0,y(0)=0.05,\, z(0)=0,  \, p_x(0)=0,\,p_y(0)=0,\,p_z(0)=0$. 
 The normalized electromagnetic field amplitude is  $a_0=436$ (each of the colliding waves has the amplitude equal to $a_0/3$),  
 the dissipation parameter is $\varepsilon_{rad}=1.2\times10^{-8}$, and the normalized critical QED field is $a_S=4\times 10^5$. 
 The electron trajectory in the $(y,\,z)$ plane plotted in Fig.\,\ref{FIG-12} a) and the trajectory in the phase $(y\,,p_z)$ plane 
 shown in Fig.\,\ref{FIG-12} b) look similar to the trajectories presented in Fig.\,\ref{FIG-2} a) and b).  
 The particle is trapped for a finite time within the EM field period performing relatively small scale oscillations. 
 Then after some time it over-leaps to the next EM field period. This is also clearly seen in Fig.\,\ref{FIG-12} c), where 
  its $y$-coordinate is plotted versus time. From the Poincar\'e sections  in Fig.\,\ref{FIG-12} d), which show
the particle positions in the phase plane $(p_z,p_y)$ at discrete times with
the time step equal to the period of the driving force, we may see that this process is stochastic. 
The particle over-leaping from one field period to another with small scale oscillations in between 
(see Figs. \ref{FIG-2},  \ref{FIG-3} and \ref{FIG-12}) may be interpreted 
in terms of L\'evy flights \cite{LEVYFLIGHTS, LEVYFLIGHTS1, LEVYFLIGHTS2, LEVYFLIGHTS3}.

\begin{figure*}
	   \begin{center}
    \includegraphics[width=12 cm]{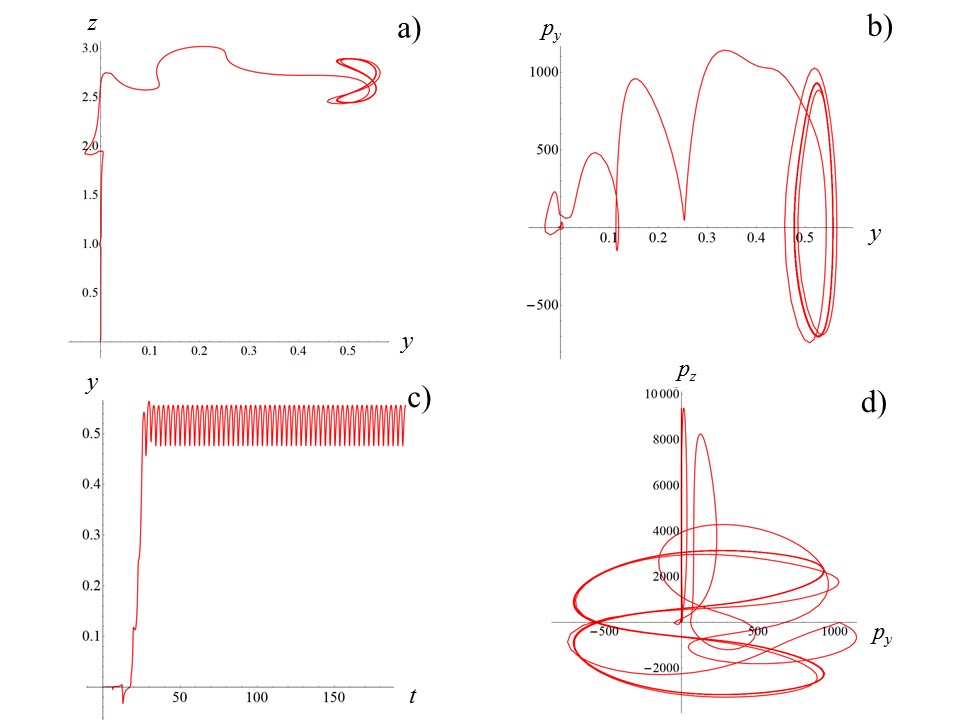}
		       \end{center}
 \caption{  a) Electron trajectory in the $(y,\,z)$ plane for initial conditions: $x(0)=0,\,y(0)=-0.0001,\, z(0)=0,
 \, p_x(0)=0,\,p_y(0)=0,\,p_z(0)=0$.
b) Trajectory in the phase  plane $(y\,,p_y)$;
c) Electron $y$-coordinate versus time;
d) Electron trajectory in the $(p_y,\,p_z)$ plane. 
The electromagnetic field amplitude is  $a_0=4700$, 
 the dissipation parameter is $\varepsilon_{rad}=1.2\times10^{-9}$, 
 and the normalized critical QED field is $a_S=4\times 10^6$. The coordinates, time and momentum
 are measured in the $2\pi c/\omega$, $2\pi/\omega$ and $m_e c$ units.
  \label{FIG-13}}
 \end{figure*}

    The case of high laser amplitude is presented in Fig.\,\ref{FIG-13}
 for initial conditions: $x(0)=0,\,y(0)=-0.0001,\, z(0)=0,  \, p_x(0)=0,\,p_y(0)=0,\,p_z(0)=0$. The normalized electromagnetic 
 field amplitude is  $a_0=4700$ (each of the colliding waves has the amplitude equal to $a_0/3$),  
 the dissipation parameter is $\varepsilon_{rad}=1.2\times10^{-9}$, and the normalized critical QED field is $a_S=4\times 10^6$. 
 The electron trajectory in the $(y,\,z)$ plane plotted in Fig.\,\ref{FIG-13} a) and the trajectory in the phase $(y\,,p_y)$ plane 
 shown in Fig.\,\ref{FIG-13} b) clearly demonstrate the particle trapping into the limit circle after an initial phase 
 corresponding to the particle motion in the vicinity of the electric field node, $y=0$. Since the motion here is unstable, 
 the particle leaves this region.   
 This is also distinctly seen in  Fig.\,\ref{FIG-13} d)  showing the electron trajectory in the $(p_y,\,p_z)$ plane. 
 In the plane $(y,\,z)$ (Fig.\,\ref{FIG-13} a)) as we see, when particle moves along the limit circle, its trajectory has 
 the ``figure eight'' form. It performs regular oscillations (see Fig.\,\ref{FIG-13} c), where 
 the particle coordinate $y$ is plotted versus time) with the double frequency for oscillations along the 
 $y$ axis compared with the frequency of oscillation  along the $z$ axis.

   \subsubsection{Random-walk and regular patterns of the particle trajectories in the field of three 3 s-polarized EM waves}
   
\begin{figure*}
	   \begin{center}
    \includegraphics[width=12 cm]{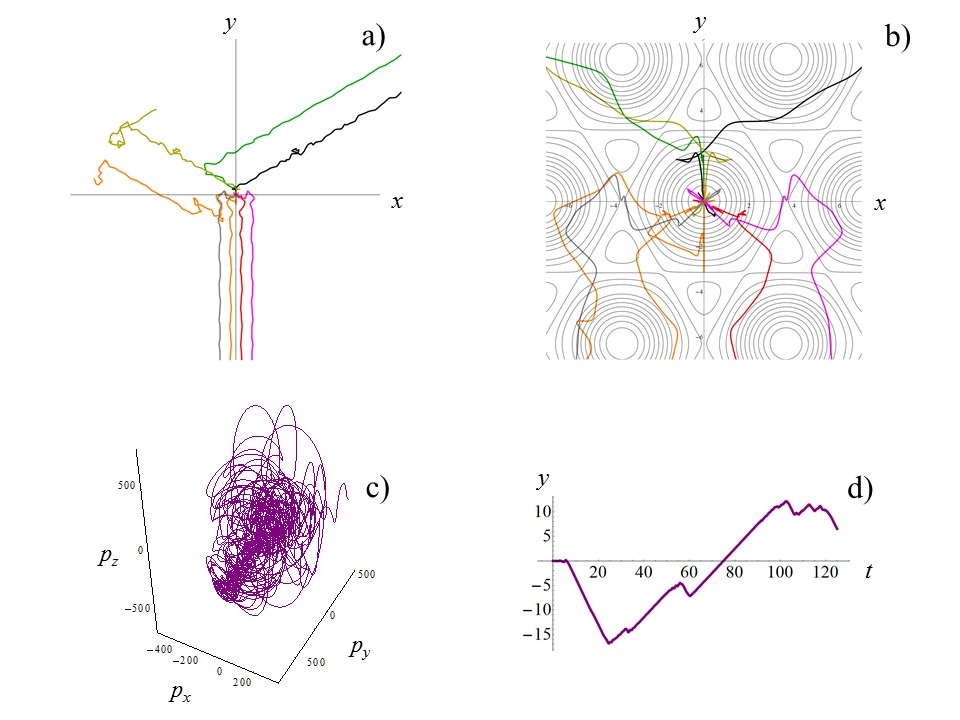}
		       \end{center}
 \caption{  a) 8 electron trajectories in the $(x,\,y)$ plane for initial conditions: $x(0)$ and $y(0)$ are in the vicinity of the coordinate origin, and $z(0)=0,
 \, p_x(0)=0,\,p_y(0)=0,\,p_z(0)=0$.
b) Close up of the trajectories in the vicinity of the coordinate origin.
c) Electron trajectory in the $p_x,\,p_y,\,p_z$ space for $x(0)=-0.125$ and $y(0)=0.125$.
d) Electron $y$ coordinate versus time for $x(0)=-0.125$ and $y(0)=0.125$. 
The electromagnetic field amplitude is  $a_0=756$, 
 the dissipation parameter is $\varepsilon_{rad}=1.2\times10^{-8}$, 
 and the normalized critical QED field is $a_S=4\times 10^5$. The coordinates, time and momentum
 are measured in the $2\pi c/\omega$, $2\pi/\omega$ and $m_e c$ units.
  \label{FIG-14}}
 \end{figure*}

Results of integrations of the motion equations for the electron interacting with  three 3 s-polarized EM waves in the limit of 
relatively low radiation intensity are presented in Fig.\,\ref{FIG-14}.
Fig.\,\ref{FIG-14}   a)  shows 8 electron trajectories in the $(x,\,y)$ plane for initial conditions: $x(0)$ and $y(0)$ are in the vicinity of the coordinate origin, and $z(0)=0,
 \, p_x(0)=0,\,p_y(0)=0,\,p_z(0)=0$. In Fig.\,\ref{FIG-14} b) we plot a close up of the trajectories in the vicinity of the coordinate origin superimposed with the isocontours of the electromagnetic potential averaged 
 over a half period of the field oscillations. It is proportional to the ponderomotive potential in the high field amplitude limit, $a_0 \gg 1$. 
 As we see the typical trajectories are comprised of  long range L\'evy-flight-like excursions and  short range rambling motion, which changes the direction of succeeding flight. The combination of  
 the long range excursions and short range rambling is also seen in the dependence of the electron $y$ coordinate on time in Fig.\,\ref{FIG-14}   d). The corresponding particle trajectory 
 in the $p_x,\,p_y,\,p_z$ momentum space for $x(0)=-0.125$ and $y(0)=0.125$ is presented in Fig.\,\ref{FIG-14} c). What is remarkable is that during the L\'evy-like flights the electron moves almost 
 along the direction of one of the three waves propagation (compare Figs.\, \ref{FIG-10} and Fig.\,\ref{FIG-14} a) ). This stage of the particle motion can be described by the particular solution analyzed above and 
 illustrated in Fig.\,\ref{FIG-12}.

\begin{figure*}
	   \begin{center}
    \includegraphics[width=12 cm]{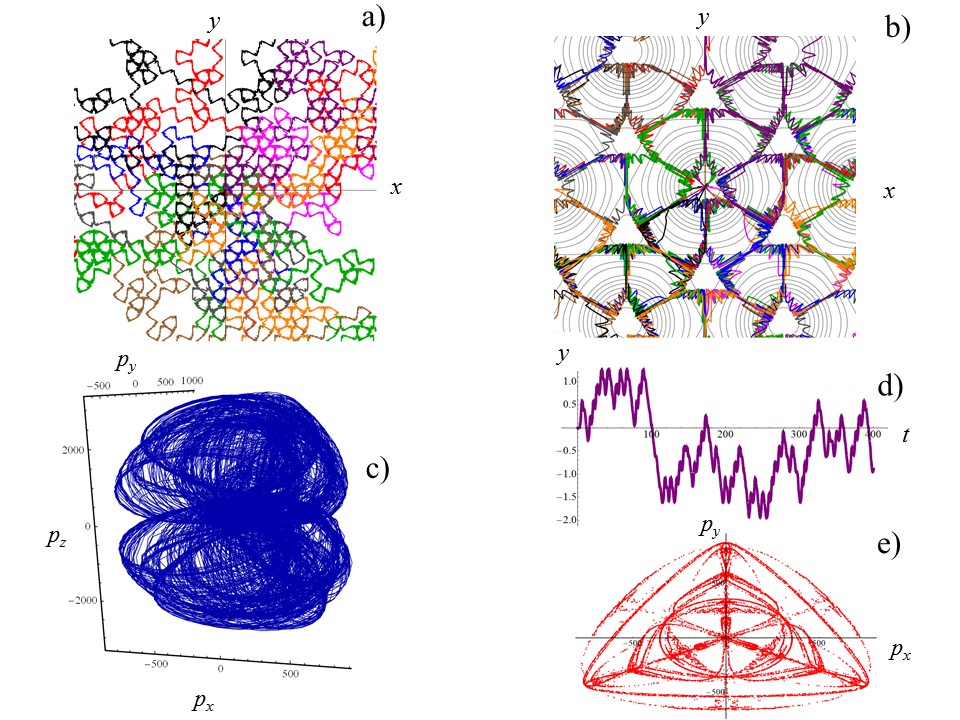}
		       \end{center}
 \caption{  a) Ensemble of electron trajectories in the $(x,\,y)$ plane for initial conditions: $x(0)$ and $y(0)$ are in the vicinity of the coordinate origin, and $z(0)=0,
 \, p_x(0)=0,\,p_y(0)=0,\,p_z(0)=0$.
b) Close up of the trajectories in the vicinity of the coordinate origin.
c) Electron trajectory in the $p_x,\,p_y,\,p_z$ space for $x(0)=-0.125$ and $y(0)=0.125$.
d) Electron $y$ coordinate versus time for $x(0)=-0.001$ and $y(0)=-0.001$. 
e) The Poinca\'re sections: the particle positions in the phase plane $(p_x,\,p_y)$ at  discrete times with
the time step equal to the period of the driving force. 
The electromagnetic field amplitude is  $a_0=4764$, 
 the dissipation parameter is $\varepsilon_{rad}=6\times10^{-9}$, 
 and the normalized critical QED field is $a_S=8\times 10^5$. The coordinates, time and momentum
 are measured in the $2\pi c/\omega$, $2\pi/\omega$ and $m_e c$ units. The integration time equals $200\times 2 \pi/ \omega$.
  \label{FIG-15}}
 \end{figure*}

 Qualitatively different patterns formed by the trajectories of particles  interacting with the field of three 3 s-polarized EM waves are observed in the high intensity and low frequency limit. 
These patterns are shown  in Fig.\,\ref{FIG-15} a) and in Fig.\,\ref{FIG-15} b) presenting a close up of the trajectories in the vicinity of the coordinate origin, where the trajectories  in the $(x,\,y)$ plane 
of an electron ensemble make a  tracery striking the imagination reminding one of a parquetry or window frost.
Either an individual trajectory or their ensemble appear to be confined in the lower measure sub-domain periodic in the $x$ and $y$ directions.
 In Fig.\,\ref{FIG-15} c) the electron trajectory in the $p_x,\,p_y,\,p_z$ momentum space for $x(0)=-0.125$ and $y(0)=0.125$ demonstrates that the particle energy remains finite.
The electron $y$ coordinate dependence on time for $x(0)=-0.001$ and $y(0)=-0.001$ plotted  in Fig.\,\ref{FIG-15} c) 
shows that the particle motion is comprised of relatively long over-leaps interlaced with  small-scale oscillations. 
In Fig.\,\ref{FIG-15} e) we present the corresponding Poinca\'re sections, i. e. we plot the particle positions in the phase plane $(p_x,\,p_y)$ at  discrete times with
the time step equal to the period of the driving force. The Poinca\'re sections, in this case, indicate that the particle motion is pretty regular.
Here the parameters of the EM field and of the electrons are as follows. The electromagnetic field amplitude is  $a_0=4764$ (the amplitude of each colliding waves is equal to 1588), 
 the dissipation parameter is $\varepsilon_{rad}=6\times10^{-9}$, 
 and the normalized critical QED field is $a_S=8\times 10^5$, which corresponds to the wave frequency a factor two smaller than in the case shown in Fig.\,\ref{FIG-14}. 
 The integration time equals $200\times 2 \pi/ \omega$

 \subsubsection{Ergodization or not?}

   \begin{figure}
	   \begin{center}
    \includegraphics[width=9 cm]{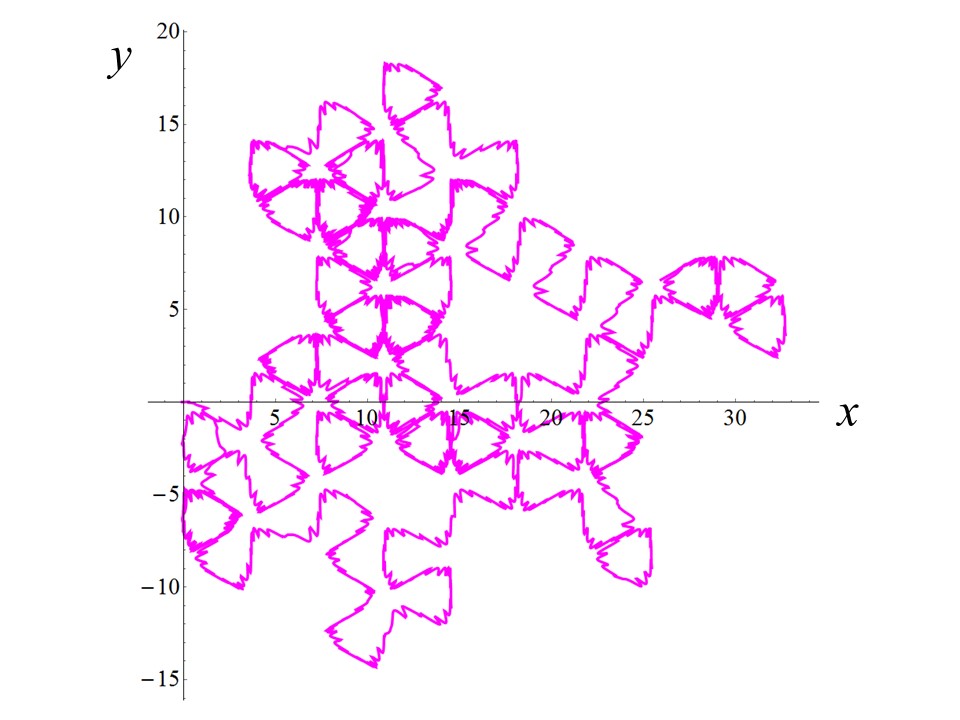}
		       \end{center}
 \caption{  Trajectory of the electron migrating over a long time in the $(x,y)$ plane.
  \label{FIG-16}}
 \end{figure}

The attractor trajectory pattern in Fig.\,\ref{FIG-15} a) and b) is made by an ensemble of electrons. 
The single electron trajectory shown in Fig.\,\ref{FIG-16} demonstrates that having been moving 
for a long enough time it could cover the whole attractor. In view of this, there are two questions.
The first one being is there an analogy of the ergodic hypothesis saying that over long periods of time, the time spent 
in some region of the attractor is proportional to the attractor measure? The second one being is there an analogy of the 
Poincar\'e recurrence theorem \cite{Arnold-1989} saying that the particle, after a sufficiently long but finite time, returns to a point very close 
to the initial point?  A similar question occurs in the case of the particle random walk 
on whether the results of well known random walk theory \cite{MKac-1947}
can 
be used in our case.
 
\bigskip 

 \subsection{Electron interaction with three p-polarized  EM waves}
   
\begin{figure*}
	   \begin{center}
    \includegraphics[width=12 cm]{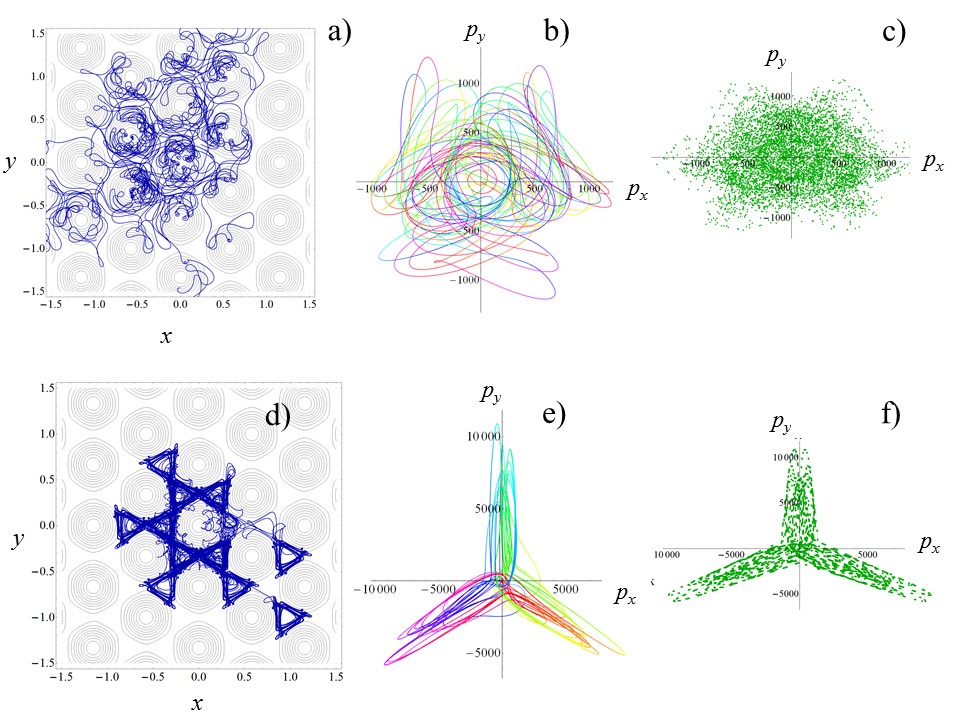}
		       \end{center}
 \caption{  a) Ensemble of electron trajectories in the $(x,\,y)$ plane for initial conditions: 
 $x(0)$ and $y(0)$ are in the vicinity of the coordinate origin, and $z(0)=0,
 \, p_x(0)=0,\,p_y(0)=0,\,p_z(0)=0$.
b) Electron trajectories in the $(p_x,\,p_y)$ plane.
c) The Poinca\'re sections: the particle positions in the phase plane $(p_x,\,p_y)$ at discrete times with
the time step equal to the period of the driving force. 
The electromagnetic field amplitude is  $a_0=1383$, 
 the dissipation parameter is $\varepsilon_{rad}=1.2\times10^{-6}$, 
 and the normalized critical QED field is $a_S=4.1\times 10^5$. 
d) Ensemble of electron trajectories in the $(x,\,y)$ plane for initial conditions: 
$x(0)$ and $y(0)$ are in the vicinity of the coordinate origin, and $z(0)=0,
 \, p_x(0)=0,\,p_y(0)=0,\,p_z(0)=0$.
e) Electron trajectories in the $(p_x,\,p_y)$ plane.
f) The Poinca\'re sections: the particle positions in the phase plane $(p_x,\,p_y)$ at discrete times with
the time step equal to the period of the driving force. 
The electromagnetic field amplitude is  $a_0=7.2\times 10^3$, 
 the dissipation parameter is $\varepsilon_{rad}=1.2\times10^{-8}$, 
 and the normalized critical QED field is $a_S=4.1\times 10^5$.
  \label{FIG-17}}
 \end{figure*}

   In the case of   three p-polarized  EM waves the EM configuration is described 
   by Eqs. (\ref{eq:EBzfield3waves}), (\ref{eq:EBfield3waves}), (\ref{eq:BEfield3waves}).
   As in the s-polarization case, in the limit of relatively low EM wave intensity  
   the electron performs the random walk motion comprised of short scale-length oscillations 
   interleaved by long scale-length L\'evy-like flights. An example of such the trajectory is shown in 
   Fig.\,\ref{FIG-17} a) for the EM field amplitude of  $a_0=4764$, 
 the dissipation parameter of $\varepsilon_{rad}=6\times10^{-9}$, 
 and the normalized critical QED field of $a_S=8\times 10^5$. 
 For the high intensity EM wave case the electrons migrate along the paths confined in narrow valleys 
 as can be seen in   Fig.\,\ref{FIG-17} b), where the ensemble of the electron trajectories  is plotted 
 for the EM field amplitude of  $7.2\times 10^3$, 
 the dissipation parameter of $\varepsilon_{rad}=1.2\times10^{-8}$, 
 and the normalized critical QED field of $a_S=4.1\times 10^5$. 

 \section{Electron dynamics in  four s- and p-polarized 
colliding EM pulses}

\begin{figure}
\centering
\includegraphics[width=5 cm]{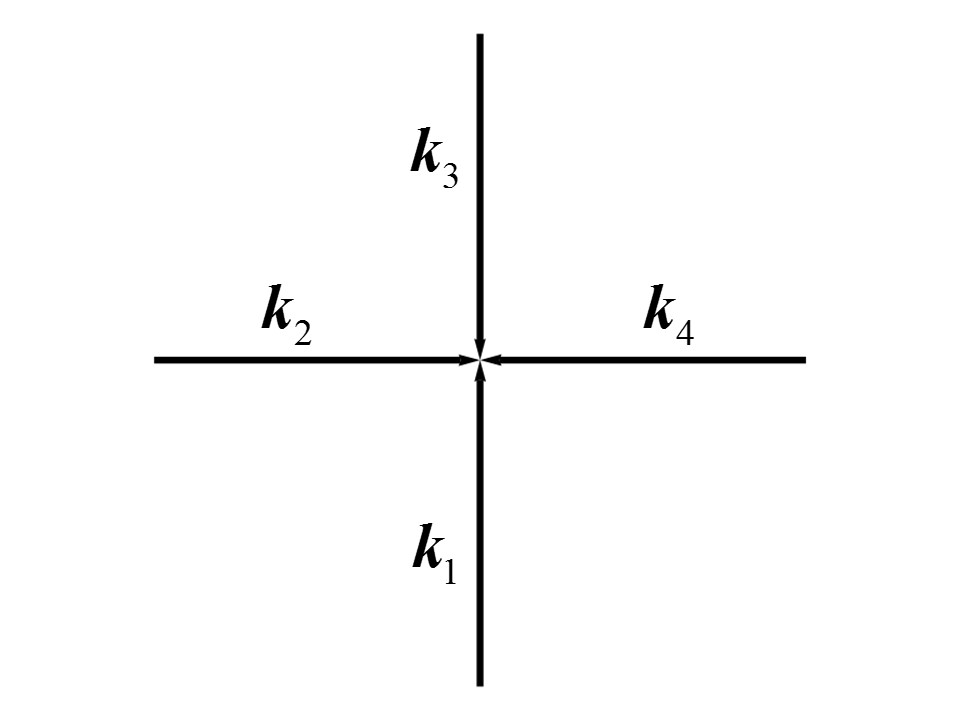}
\caption{ Wave vectors of four colliding EM waves.}
\label{FIG-18}
\end{figure}

\begin{figure}
\includegraphics[width=8 cm]{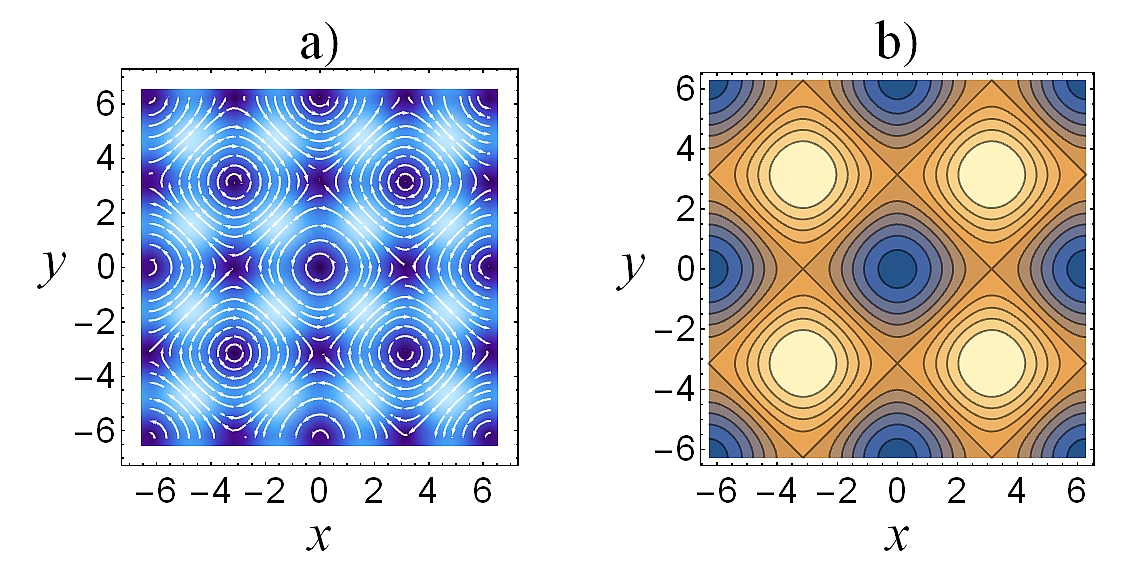}
\caption{ Four s-polarized (p-polarized) EM waves: a) magnetic (electric) field; 
b) isocontours of the electric (magnetic) field in the $(x,y)$ plane at time $t=\pi/4$.}
\label{FIG-19}
\end{figure}

The  orientation of four colliding waves is illustrated in Fig.\,\ref{FIG-18}.
 Fig.\,\ref{FIG-19} a)  shows magnetic (electric) field and  
b) isocontours of the electric (magnetic) field in the $(x,y)$ plane at time $t=\pi/4$ 
of four s-polarized (p-polarized) colliding EM waves.

\subsection{S-polarized 4 colliding EM waves}

\subsection{EM field configuration}

In the EM configuration of four colliding s(p)-polarized waves 
the  $z$-components of the electric (magnetic) field can be written 
\begin{equation}
\left(
 \begin{array}{c}
   E_z \\
   B_z
 \end{array}
\right)
=
\left(
 \begin{array}{c}
   E_0 \\
   B_0
 \end{array}
\right)2\sin(\omega_0 t) \left[\cos\left (\omega_0 \frac{x}{c}\right)+\cos\left (\omega_0 \frac{y}{c}\right)\right].
\label{eq:EBzfield4waves}
\end{equation}
The $x$ and $y$ components of the magnetic (electric) field of the four colliding s(p)-polarized waves are given by
\begin{equation}
\left(
 \begin{array}{c}
   B_x \\
   E_x 
 \end{array}
\right)
=
\left(
 \begin{array}{r}
   -E_0 \\
  B_0 
 \end{array}
\right)
2\cos(\omega_0 t)\sin\left(\omega_0\frac{y}{c}\right)
\label{eq:EBfield4waves}
\end{equation}
and
\begin{equation}
\left(
 \begin{array}{c}
   B_y \\
   E_y 
 \end{array}
\right)
=
\left(
 \begin{array}{r}
   E_0 \\
  -B_0 
 \end{array}
\right)2\cos(\omega_0 t)\sin\left(\omega_0\frac{x}{c}\right),
\label{eq:BEfield4waves}
\end{equation}
respectively.

\subsubsection{Particular solutions}

As in the above considered case of three s-polarized EM waves the equations of electron motion admit 
particular solutions, in the first of which the particle moves either along one of the axis, i.e. 
$x=n c\pi/\omega$ or  $y=n c\pi/\omega$ with $n=0,\pm 1, \pm2, \,...$, and in the second 
it moves along  straight lines $x=\pm y+n c\pi/\omega$.

\begin{figure*}
\centering
\includegraphics[width=12 cm]{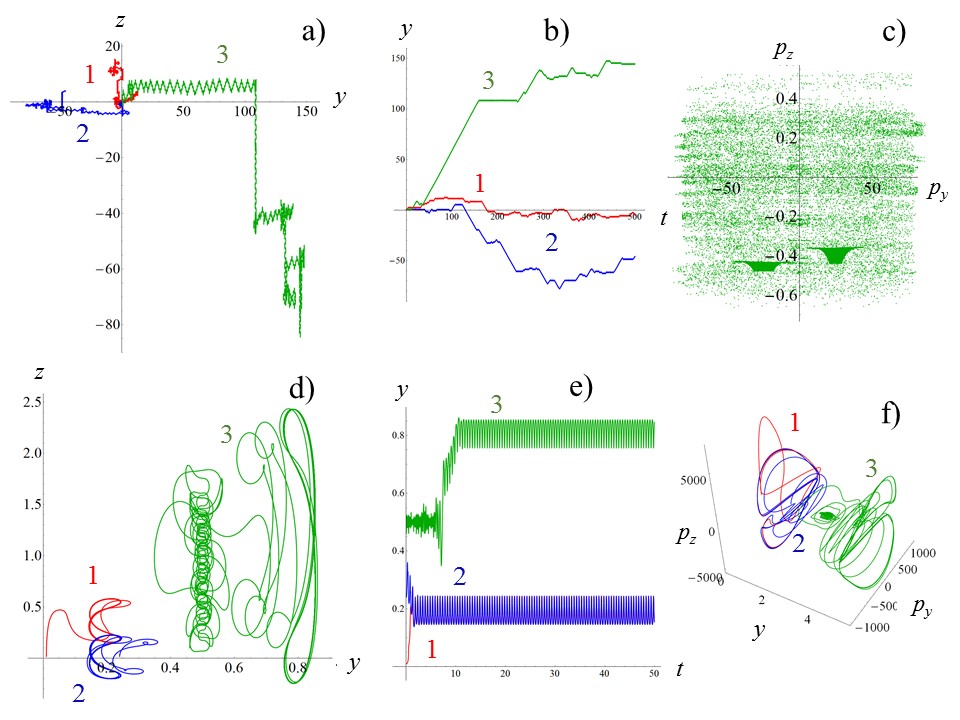}
\caption{  Electron trajectories in the case of {\bf the first type} particular solution corresponding to the 
motion along the $y$ axis (at $x=0$) in the field of four colliding EM waves 
for $\varepsilon_{rad}=1.2\times 10^{-9}$, $a_S=4\times 10^6$, $a=94$, $\omega=0.1$
 for initial conditions: $y(0)=0.01,\, z(0)=0, \, p_y(0)=0,\,p_z(0)=0$ (red,-1); 
$y(0)=0.23,\, z(0)=0, \, p_y(0)=0,\,p_z(0)=0$ (blue,-2); $y(0)=0.45,\, z(0)=0, \, p_x(0)=0,\,p_z(0)=0$ (green,-3).
a) Trajectories in the $(y,z)$ plane. 
b) Dependences of the $y$ coordinates on time.
c)  The Poinca\'re sections: the particle positions in the phase plane $(p_y,\,p_z)$ at discrete times with
the time step equal to the period of the driving force.
For lower frequency,  $\omega=0.02$, when  $\varepsilon_{rad}=2.4\times 10^{-10}$, $a_S=2\times 10^7$, $a_0=8\times 10^3$, they are shown 
d) trajectories in the $(y,\,z)$ plane, e)  dependences of the $y$ coordinates on time, and f) trajectories in the $(y,p_y,p_z)$ space 
for the same initial conditions as in the frames a,b,c).}
\label{FIG-20}
\end{figure*}

{\bf First type solution.} For the first class of particular solutions with $x= n c\pi/\omega$ (without loss of generality we may take $n=0$, i. e. consider $x=0$), 
formally the particle moves in a superposition of the fields of two counter-propagating s-polarized EM waves 
and a homogeneous oscillating electric field directed along the $z$ axis. As in the above considered cases of two 
and three colliding EM waves, in the limit of weak nonlinearity and dissipation 
($\varepsilon_{rad}=1.2\times 10^{-9}$, $a_S=4\times 10^6$, $a=94$, $\omega=0.1$, for initial conditions: $y(0)=0.01,\, z(0)=0, \, p_x(0)=0,\,p_z(0)=0$ (red,-1); 
$y(0)=0.23,\, z(0)=0, \, p_x(0)=0,\,p_z(0)=0$ (blue,-2); $y(0)=0.45,\, z(0)=0, \, p_x(0)=0,\,p_z(0)=0$ (green,-3)) 
the particle motion can be described as a 
random walk, for which the trajectories consist of the relatively small amplitude fast oscillating parts and of the 
long scale length L\'evy flights (see Fig.\,\ref{FIG-20} a) and b)).  The Poinca\'re sections, the particle positions 
in the phase plane $(p_y,\,p_z)$ at discrete times with
the time step equal to the period of the driving force, presented in Fig.\,\ref{FIG-20} c)  show that the electron motion is stochastic.

In the case of lower frequency, $\omega=0.02$, and higher dimensionless EM field amplitude  $a=8\times 10^3$, when  
$\varepsilon_{rad}=2.4\times 10^{-10}$, $a_S=2\times 10^7$, $a_0=8\times 10^3$, $\omega=0.02$, the electron trajectories 
in the  the $(y,z)$ plane (see Fig.\,\ref{FIG-20} d)) show that the particles are trapped within narrow regions moving along 
regular limit circles  (Fig.\,\ref{FIG-20} e)). The attractor geometry is distinctly seen in Fig.\,\ref{FIG-20} f), 
where the trajectories in the $(y,\,p_y,\,p_z)$ space are presented. As well seen, after a relatively short initial time interval the 
particles are trapped into stable limit circles  performing periodic motion.
We note that for the parameters chosen although the particle energy is ultrarelativistic the value of $\chi_e$ remains below 
unity, i.e. the QED effect of the recoil is not significant. 

\begin{figure*}
\centering
\includegraphics[width=12 cm]{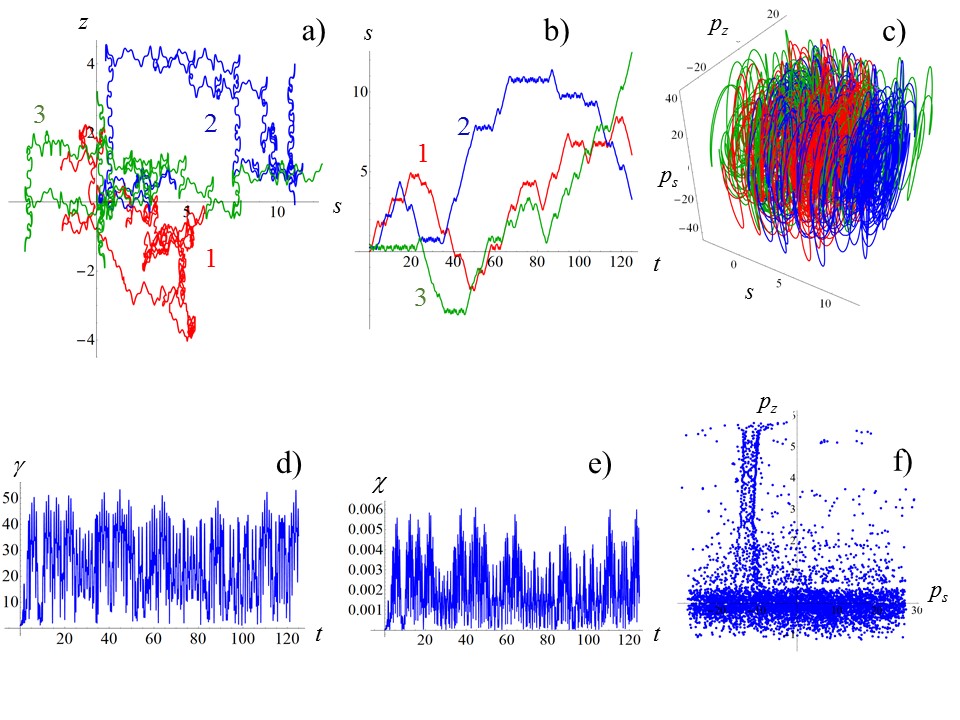}
\caption{  Electron trajectories in the case of {\bf the second type} particular solution corresponding to the 
motion along the $x=y$ direction in the field of four colliding EM waves 
for $\varepsilon_{rad}=1.2\times 10^{-8}$, $a_S=4\times 10^5$, $a_0=44$, $\omega=1$
 for initial conditions: $x(0)=0.01,\, z(0)=0, \, p_x(0)=0,\,p_z(0)=0$ (red,-1); 
$x(0)=0.23,\, z(0)=0, \, p_x(0)=0,\,p_z(0)=0$ (blue,-2); $x(0)=0.45,\, z(0)=0, \, p_x(0)=0,\,p_z(0)=0$ (green,-3).
a) Trajectories in the $(s,\,z)$ plane. 
b) Dependences of the $s$ coordinates on time.
c)  Trajectories in the $(s,\,p_s,\,p_z)$ space.
d) The particle Lorentz factor $\gamma$  versus time.
e) Parameter $\chi$ versus time.
f)  The Poinca\'re sections: the particle positions in the phase plane $(p_s,\,p_z)$ at discrete times with
the time step equal to the period of the driving force.
}
\label{FIG-21}
\end{figure*}

\begin{figure*}
\centering
\includegraphics[width=12 cm]{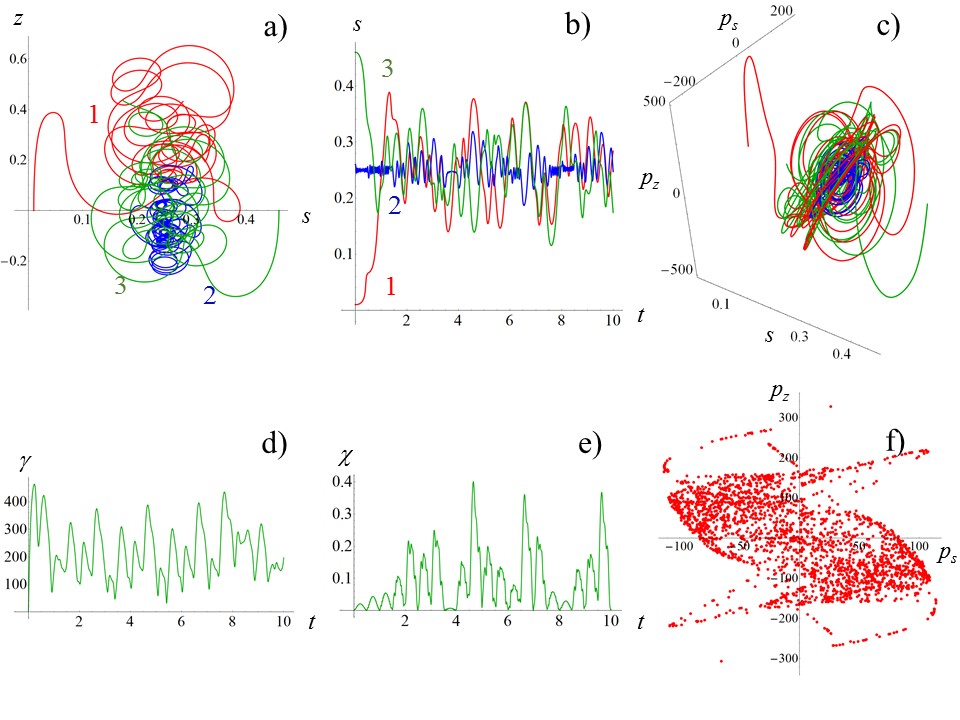}
\caption{ Electron trajectories in the case of {\bf the second type} particular solution corresponding to the 
motion along the $x=y$ direction in the field of four colliding EM waves 
for $\varepsilon_{rad}=1.2\times 10^{-8}$, $a_S=4\times 10^5$, $a_0=874$, $\omega=1$
 for initial conditions: $x(0)=0.01,\, z(0)=0, \, p_x(0)=0,\,p_z(0)=0$ (red,-1); 
$x(0)=0.23,\, z(0)=0, \, p_x(0)=0,\,p_z(0)=0$ (blue,-2); $x(0)=0.45,\, z(0)=0, \, p_x(0)=0,\,p_z(0)=0$ (green,-3).
a) Trajectories in the $(s,\,z)$ plane. 
b) Dependences of the $s$ coordinates on time.
c) Trajectories in the $(s,\,p_s\,,p_z)$ space. 
d) Lorentz factor $\gamma$ versus time for $x(0)=0.45$.
e) Parameter $\chi_e$ versus time for $x(0)=0.45$.
f)  The Poinca\'re sections: the particle positions in the phase plane $(p_s,\,p_z)$ at discrete times with
the time step equal to the period of the driving force for $x(0)=0.01$.}
\label{FIG-22}
\end{figure*}

{\bf Second type solution.} The particle behavior under the conditions corresponding to the second class of particular solutions of the equations of motion 
($x=y$) is illustrated in Figs.\,\ref{FIG-21} and \ref{FIG-22}. Here the coordinate $s(t)$ is equal to $s=x=y$.

In Fig. \,\ref{FIG-21} we present electron trajectories in the case  corresponding to the 
motion along the $x=y$ direction in the field of four colliding EM waves 
for $\varepsilon_{rad}=1.2\times 10^{-8}$, $a_S=4\times 10^5$, $a_0=44$, $\omega=1$
 for initial conditions: $x(0)=0.01,\, z(0)=0, \, p_x(0)=0,\,p_z(0)=0$ (red,-1); 
$x(0)=0.23,\, z(0)=0, \, p_x(0)=0,\,p_z(0)=0$ (blue,-2); $x(0)=0.45,\, z(0)=0, \, p_x(0)=0,\,p_z(0)=0$ (green,-3).
Fig. \,\ref{FIG-21} a) shows electron trajectories in the $(s,\,z)$ plane, which demonstrate random walks with 
intermittent short scale length oscillations and long range L\'evy flights. The same behavior is distinctly seen in 
Fig. \,\ref{FIG-21} b) with three dependences of the $s$ coordinates on time. Stochastic character of the 
particle motion is demonstrated in Figs. \,\ref{FIG-21} b) and f) by the behavior of trajectories in the $(s,\,p_s,\,p_z)$ space
and by  the particle positions in the phase plane $(p_s,\,p_z)$ at discrete times with
the time step equal to the period of the driving force, respectively. 
According to Fig. \,\ref{FIG-21} d), where the particle Lorentz factor $\gamma$  is plotted versus time, 
the normalized electron energy is of the order of the dimensionless EM field amplitude, i. e. $\gamma \approx a_0$.
From  the dependence of the parameter $\chi$ on time in Fig. \,\ref{FIG-21} e) it follows that, in this case, 
the QED effect of the recoil is not significant.

\begin{figure*}
\centering
\includegraphics[width=12 cm]{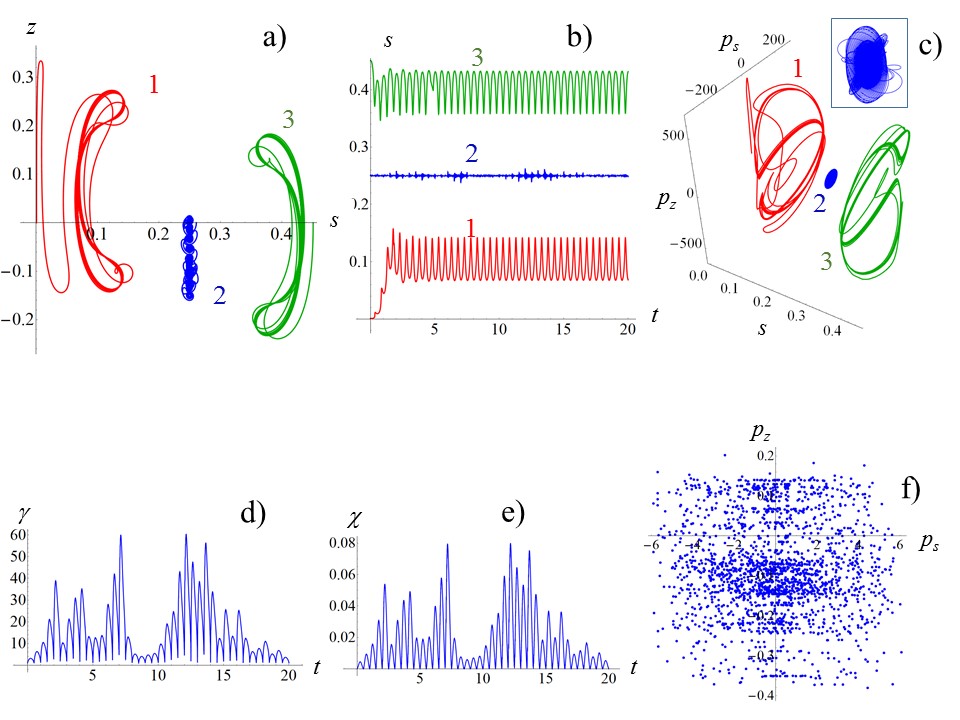}
\caption{ Electron trajectories in the case of {\bf the second type} particular solution corresponding to the 
motion along the $x=y$ direction in the field of four colliding EM waves 
for $\varepsilon_{rad}=3\times 10^{-9}$, $a_S=1.6\times 10^6$, $a_0=3466$, $\omega=0.25$
 for initial conditions: $x(0)=0.01,\, z(0)=0, \, p_x(0)=0,\,p_z(0)=0$ (red,-1); 
$x(0)=0.23,\, z(0)=0, \, p_x(0)=0,\,p_z(0)=0$ (blue,-2); $x(0)=0.45,\, z(0)=0, \, p_x(0)=0,\,p_z(0)=0$ (green,-3).
a) Trajectories in the $(s,\,z)$ plane. 
b) Dependences of the $s$ coordinates on time.
c) Trajectories in the $(s,\,p_s\,,p_z)$ space. The inset shows a close-up of trajectories in the $(s,\,p_s\,,p_z)$ for $x(0)=0.23$.
d) Lorentz factor $\gamma$ versus time for $x(0)=0.23$.
e) Parameter $\chi_e$ versus time for $x(0)=0.23$.
f)  The Poinca\'re sections: the particle positions in the phase plane $(p_s,\,p_z)$ at discrete times with
the time step equal to the period of the driving force for $x(0)=0.23$.}
\label{FIG-23}
\end{figure*}

The electron interaction with four colliding EM waves in the case of {\bf the second type} particular solution corresponding to the 
motion along the $x=y$ direction is illustrated in Fig. \,\ref{FIG-22}   
for $\varepsilon_{rad}=1.2\times 10^{-8}$, $a_S=4\times 10^5$, $a_0=874$, $\omega=1$
 for initial conditions: $x(0)=0.01,\, z(0)=0, \, p_x(0)=0,\,p_z(0)=0$ (red,-1); 
$x(0)=0.23,\, z(0)=0, \, p_x(0)=0,\,p_z(0)=0$ (blue,-2); $x(0)=0.45,\, z(0)=0, \, p_x(0)=0,\,p_z(0)=0$ (green,-3).
The particle independently of the initial conditions becomes trapped by a strange attractor performing stochastic motion.
Frame Fig. \,\ref{FIG-22} a) shows trajectories in the $(s,\,z)$ plane. As we see 
the electrons become trapped  in the region of the ponderomotive force  minimum. 
From Figs. \,\ref{FIG-23} b) and c) with dependences of the $s$ coordinates on time and with the trajectories in the $(s,\,p_s\,,p_z)$ space 
it follows that the trapped particle motion with all the three initial conditions is irregular. 
As we may see in Fig. \,\ref{FIG-22} d),  where the Lorentz factor $\gamma$ versus 
time for $x(0)=0.23$ is presented,  the normalized particle 
 energy is of the order of the dimensionless EM wave amplitude. 
 The QED parameter $\chi_e$, whose dependence on time for $x(0)=0.23$ 
 is shown in Fig. \,\ref{FIG-22} e) is  lower than unity,
  i. e. the QED effect of the recoil is  weak.
  The Poinca\'re sections are shown  in Fig. \,\ref{FIG-22} e): the particle positions in the phase plane $(p_s,\,p_z)$ at discrete times with
the time step equal to the period of the driving force for $x(0)=0.01$. As we see, 
the particle motion is stochastic.

Electron interaction with four colliding EM waves in the case of {\bf the second type} particular solution corresponding to the 
motion along the $x=y$ direction is illustrated in Fig. \,\ref{FIG-23}   
for $\varepsilon_{rad}=3\times 10^{-9}$, $a_S=1.6\times 10^6$, $a_0=3466$, $\omega=0.25$
 for initial conditions: $x(0)=0.01,\, z(0)=0, \, p_x(0)=0,\,p_z(0)=0$ (red,-1); 
$x(0)=0.23,\, z(0)=0, \, p_x(0)=0,\,p_z(0)=0$ (blue,-2); $x(0)=0.45,\, z(0)=0, \, p_x(0)=0,\,p_z(0)=0$ (green,-3).
In this case the EM wave frequency is lower than in the above discussed case and the EM wave amplitude is higher.
As a result the particle is trapped performing either regular or stochastic motion.
Frame Fig. \,\ref{FIG-23} a) shows trajectories in the $(s,\,z)$ plane. As we see depending on the initial conditions 
the electron becomes trapped either in the region of the ponderomotive force maximum or in the region of its minimum. 
From Figs. \,\ref{FIG-23} b) and c) with dependences of the $s$ coordinates on time and with the trajectories in the $(s,\,p_s\,,p_z)$ space it 
follows that the trapped particle motion with the initial conditions $x(0)=0.01$ and  $x(0)=0.45$ along the limit circles is regular. 
As one can see in Fig. \,\ref{FIG-23} d),  where the Lorentz factor $\gamma$ versus 
time for $x(0)=0.23$ is presented,  the normalized particle 
 energy is substantially lower than the dimensionless EM wave amplitude. 
 The QED parameter $\chi_e$, whose dependence on time for $x(0)=0.23$ 
 is shown Fig. \,\ref{FIG-23} e) is significantly lower than unity,
  i. e. the QED effect of the recoil is negligibly weak.
  The Poinca\'re sections are shown  in Fig. \,\ref{FIG-23} e): the particle positions in the phase plane $(p_s,\,p_z)$ at discrete times with
the time step equal to the period of the driving force for $x(0)=0.23$. As we see, 
the particle motion along the trajectories of the attractor plotted in the inset in Fig. \,\ref{FIG-23} c) 
with the close-up of trajectories in the $(s,\,p_s\,,p_z)$ for $x(0)=0.23$ is stochastic.

\subsubsection{General case}

\begin{figure*}
\centering
\includegraphics[width=12 cm]{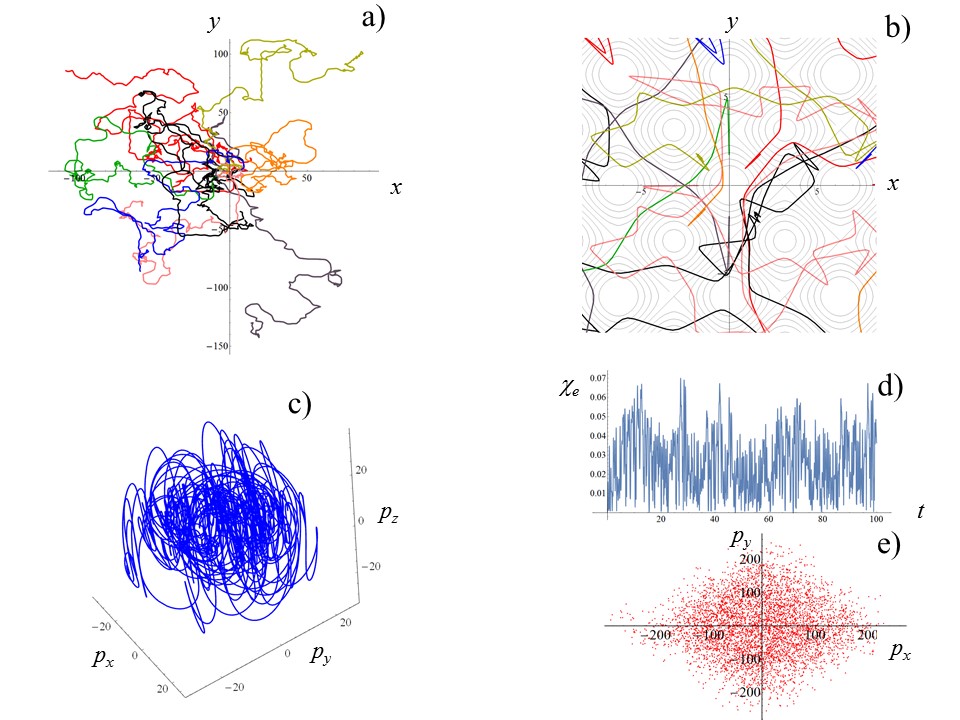}
\caption{  a) 11 electron trajectories in the $(x,y)$ plane for initial conditions: $x(0)$ and $y(0)$ are in the vicinity of the coordinate origin, and $z(0)=0$, $p_x(0)=0$, $p_y(0)=0$, $p_z(0)=0$.
b) Close up of the trajectories in the region $(-7.5<x<7.5;-7.5<y<7.5)$.
c)  Trajectory in the $(p_x,\,p_y\,,p_z)$ space. 
d) Parameter $\chi_e$ versus time.
e)  The Poinca\'re sections: the particle positions in the phase plane $(p_x,\,p_y)$ at discrete times with
the time step equal to the period of the driving force.
The electromagnetic field amplitude is $a_0=218$, the dissipation parameter is $\varepsilon_{rad}=1.2\times10^{-8}$, 
the normalized critical QED field is $a_S=4\times10^5$,  and the EM field frequency equals $\omega_0=1$. }
\label{FIG-24}
\end{figure*}

The results of integration of the motion equations for the electron interacting with four s-polarized EM waves in the limit of relatively low radiation intensity are presented in Fig.\,\ref{FIG-24}. 
Fig.\,\ref{FIG-24} a) shows 11 electron trajectories in the $(x,y)$ plane for initial conditions as follows. The  initial coordinates $x(0)$ and $y(0)$ are chosen to be in  the vicinity 
of the coordinate origin, and $z(0)=0$, $p_x(0)=0$, $p_y(0)=0$, $p_z(0)=0$. In Fig.\,\ref{FIG-24} b) we show a close up of the trajectories in the vicinity of the coordinate 
origin superimposed with the isocontours of the electromagnetic potential averaged over a half period of the field oscillations. 
It is proportional to the ponderomotive potential in the high field amplitude limit, $a_0\gg1$.  As we see, the typical trajectories are comprised of  long range L\'evy-flight-like excursions 
and of  short range rambling motion, which changes the direction of the succeeding flight. 
 Corresponding particle trajectory in the $(p_x, p_y, p_z)$ momentum space for $x(0)=-0.125$ and $y(0)=0.125$ is presented Fig.\,\ref{FIG-24} c). 
 According to the dependence of the parameter $\chi_e$ on time plotted in Fig.\,\ref{FIG-24} d) the QED recoil effects are weak under the conditions of consideration.
The Poinca\'re sections, the particle positions in the phase plane $(p_x,\,p_y)$ at discrete times with
the time step equal to the period of the driving force, in Fig.\,\ref{FIG-24} e), show that the particle motion is stochastic.

\begin{figure*}
\centering
\includegraphics[width=12 cm]{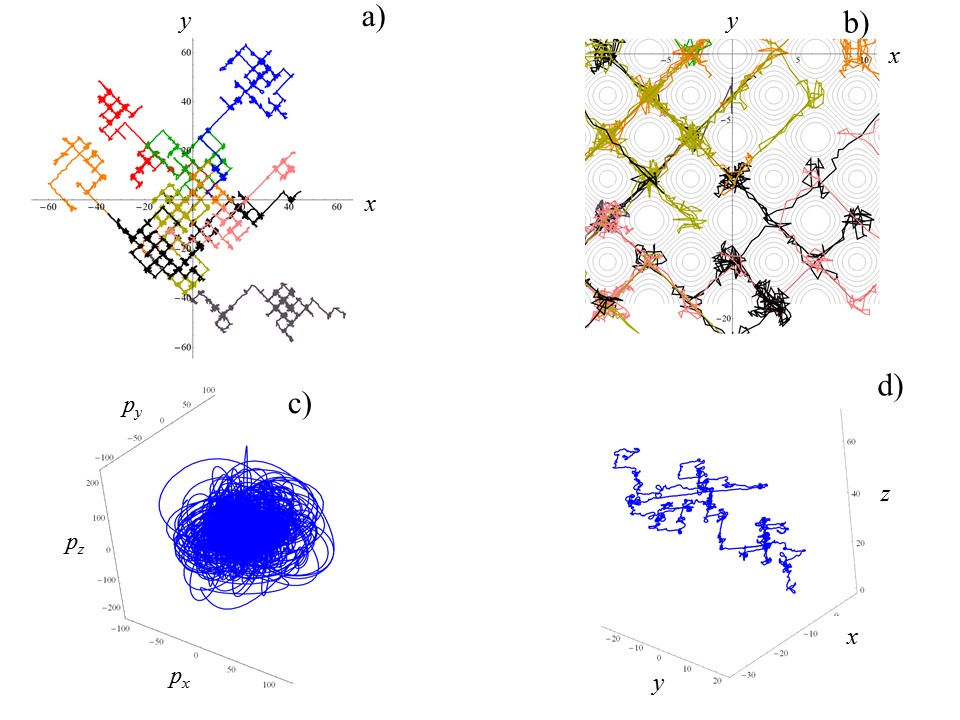}
\caption{  a) 11 electron trajectories in the $(x,y)$ plane for initial conditions: $x(0)$ and $y(0)$ 
are in the vicinity of the coordinate origin, and $z(0)=0$, $p_x(0)=0$, $p_y(0)=0$, $p_z(0)=0$.
b) Close up of the trajectories in the region $(-10<x<10;-20<y<0)$ superimposed with the isocontours 
if the electromagnetic potential averaged over a half period of the field oscillations.
c)  Trajectory in the $(p_x,\,p_y\,,p_z)$ space. 
d) Trajectory in the $(x,\,y\,,z)$ space. 
The electromagnetic field amplitude is $a_0=2823$, the dissipation parameter is $\varepsilon_{rad}=1.2\times10^{-9}$, 
the normalized critical QED field is $a_S=4\times10^6$,  and the EM field frequency equals $\omega_0=0.1$. }
\label{FIG-25}
\end{figure*}

Fig.\,\ref{FIG-25} illustrates the particle dynamics in the EM field formed by four s-polarized EM waves
 for the radiation intensity higher than that intensity which corresponds to the interaction regime shown in Fig.\,\ref{FIG-24}.
Here the electromagnetic field amplitude is $a_0=2823$, the dissipation parameter is $\varepsilon_{rad}=1.2\times10^{-9}$, 
the normalized critical QED field is $a_S=4\times10^6$,  and the EM field frequency equals $\omega_0=0.1$. 
From Fig.\,\ref{FIG-25} a) and b) it follows that the the typical trajectories form a pretty regular pattern in the $(x,y)$ plane. 
They are comprised of  long range L\'evy-flight-like excursions and of  short range rambling motion, which changes the direction of the succeeding flight. 
The combination of the long range excursions and short range rambling is also seen in the behavior of the electron trajectory in the $(x,y,z)$ space presented in Fig.\,\ref{FIG-25} d).
 The corresponding particle trajectory in the $(p_x, p_y, p_z)$ momentum space for $x(0)=-0.125$ and $y(0)=0.125$ is presented in Fig.\,\ref{FIG-25} c). 
 What is remarkable is that during the L\'evy like flights the electron moves almost along the electric node region, 
 i.e. performing the motion described by the second type particular solution 
 discussed above (see Figs.\,\ref{FIG-22}). The particle normalized energy changes from 200 to approximately 1200. 
 The value of the QED dimensionless parameter $\chi_e$ (not shown here) is less than unity.
 The  Poinca\'re sections (also not shown here) are similar to those  sections which are presented in Fig.\,\ref{FIG-24} e) 
 indicating stochasticity in the electron dynamics.
 
\begin{figure}
\centering
\includegraphics[width=6 cm]{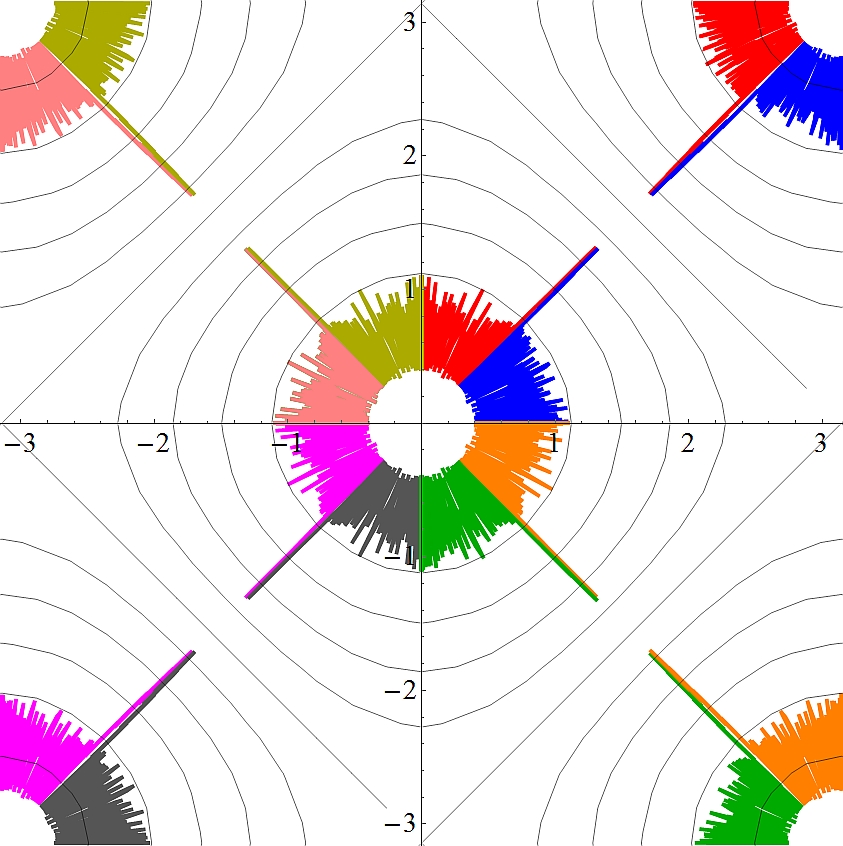}
\caption{ Ensemble of the electron trajectories in the $(x,y)$ plane. 
The particles with the initial coordinates in the region close to the $B=0$ point are trapped inside the sectors, 
where their trajectories asymptotically approach the lines $x=0$ or $y=0$. For the initial coordinates close to the bottoms of the 
ponderomotive potential valleys, $x=\pm y= \pi n, \,\,\, n=...\, , -2,-1,0,1,2, \, ...$ the particles move 
along the trajectories which are similar to those shown in Figs.\,\ref{FIG-25} a) and b). The EM 
field amplitude is $a_0=11856$, the dissipation parameter is  $\varepsilon_{rad}=6\times10^{-10}$, 
the normalized critical QED field is $a_S=8\times10^6$,  and the EM field frequency equals $\omega_0=0.05$.}
\label{FIG-26}
\end{figure}

\begin{figure*}
\centering
\includegraphics[width=12 cm]{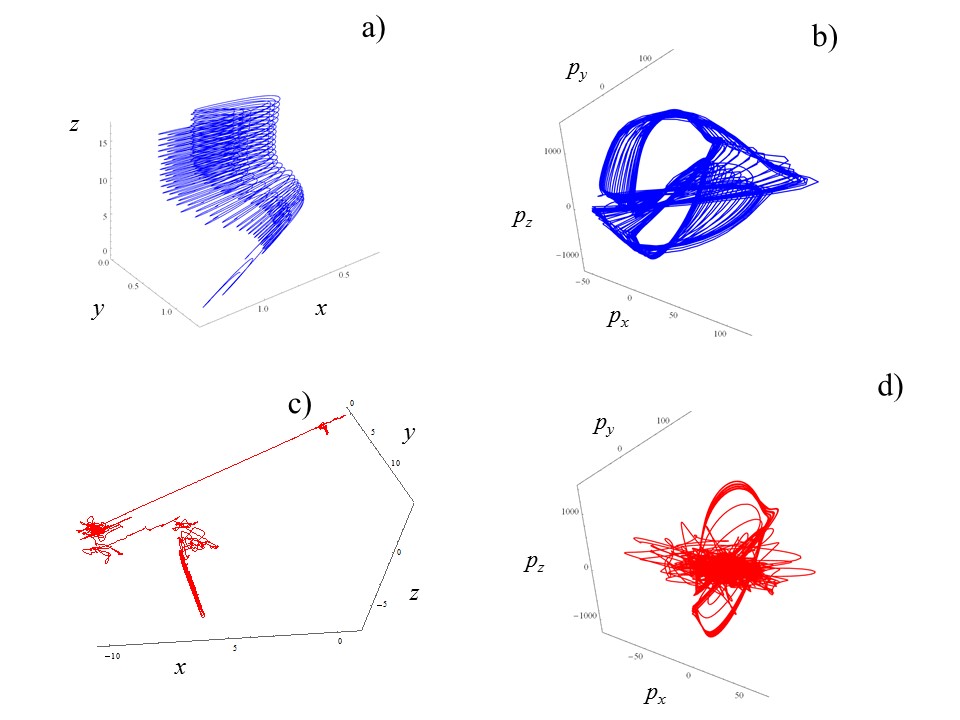}
\caption{  a) Electron trajectory in the $(x,y,z)$ space and b) trajectory in the $(p_x,p_y,p_z)$ space for the {\bf second class topology}.
c) Electron trajectory in the $(x,y,z)$ space and d) trajectory in the $(p_x,p_y,p_z)$ space for the {\bf first class topology}.
The electromagnetic field parameters are the same as in Fig.\ref{FIG-26}. }
\label{FIG-27}
\end{figure*}

Further increasing the EM field intensity and/or decreasing the field frequency lead to an intriguing change in the trajectory pattern 
(see Fig.\,\ref{FIG-26}, where an ensemble of the electron trajectories in the $(x,y)$ plane is presented). 
The results presented in  Figs.\,\ref{FIG-26} and \,\ref{FIG-27} have been obtained for the electromagnetic 
field amplitude of $a_0=11856$, for the dissipation parameter of  $\varepsilon_{rad}=6\times10^{-10}$, for
the normalized critical QED field of $a_S=8\times10^6$,  and for the EM field frequency equal to $\omega_0=0.05$.
The trajectory topology 
can be subdivided into two classes depending on the particle initial conditions. If the particle is initially close to the bottom 
of the ponderomotive potential, i.e. close to the lines $x=\pm y= \pi n, \,\,\, n=...\, , -2,-1,0,1,2, \, ...$ in the  $(x,y)$ plane, 
it remains there. The particle trajectory, in this case, is similar to those shown in Figs.\,\ref{FIG-25} a) and b). The second class trajectories 
are realized for the initial 
particle positions in the vicinity of the ponderomotive potential maximum, where the magnetic field of the colliding EM waves 
vanishes. The second class trajectories are trapped within one of the sectors, $0<\theta<\pi/4$,  $\pi/4<\theta<\pi/2$, etc.
Oscillating along the radial direction they drift  relatively slowly towards the lines either $x=0$ or $y=0$. In both the cases 
of the first and second topology classes the particles move also along the $z$ axis as seen from the results presented in Fig.  \,\ref{FIG-27}.
The first class particle dynamics is stochastic: the trajectory in the $(p_x,p_y,p_z)$ space plotted in Fig.  \,\ref{FIG-27} d)  corresponds to a  
strange attractor while  Fig.  \,\ref{FIG-27} b) shows that the second class dynamics is regular.

 \subsection{Electron interaction with four p-polarized  EM waves}
   
\begin{figure*}
	   \begin{center}
    \includegraphics[width=12 cm]{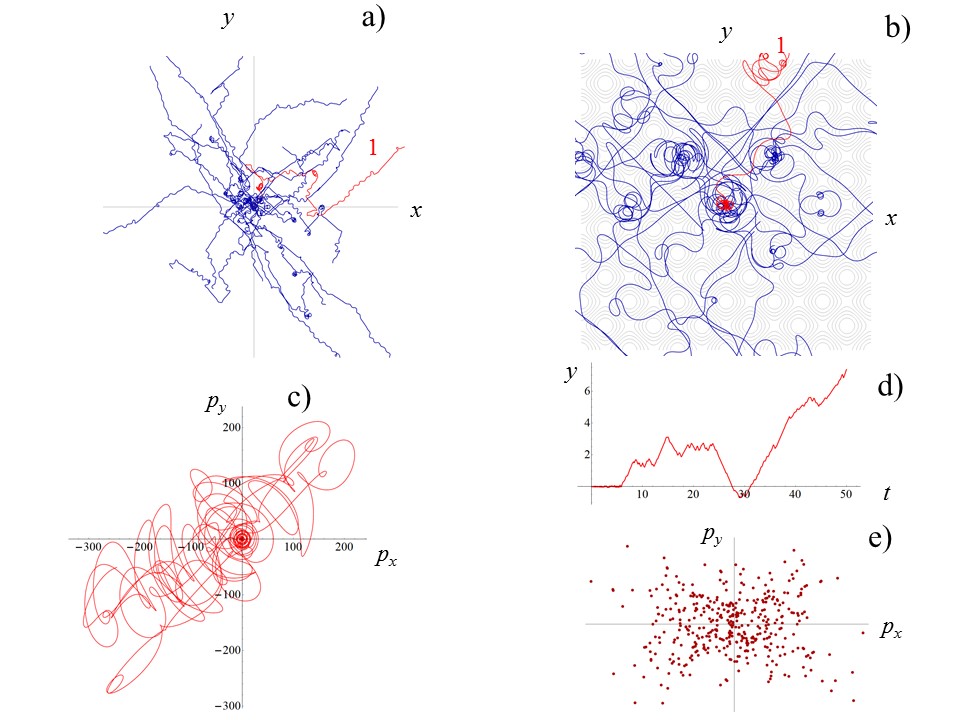}
		       \end{center}
 \caption{Electron interaction with 4 colliding p-polarized EM waves in the low intensity limit for
 the electromagnetic field amplitude equal to  $a_0=1.6\times 10^3$, 
 the dissipation parameter equal to $\varepsilon_{rad}=1.2\times10^{-8}$, 
 and the normalized critical QED field of $a_S=4.12\times 10^5$.
   a) Ensemble of electron trajectories in the $(x,\,y)$ plane. Red color (1) curve 
   corresponds to $x(0)=0.001$ and $y(0)=0.01$.
   b) Close-up of trajectory (1) the $(x,\,y)$ plane overlaid with the isocontours of the EM field ponderomotive potential. 
c) Electron trajectory in the $(p_x,\,p_y)$ plane.
d) Coordinate $y$ versus time $t$.
e) The Poinca\'re sections: the particle positions in the phase plane $(p_x,\,p_y)$ at discrete times with
the time step equal to the period of the driving force. 
  \label{FIG-28}}
 \end{figure*}

   In the case of   four p-polarized  colliding  laser pulses the EM configuration is described 
   by Eqs. (\ref{eq:EBzfield4waves}), (\ref{eq:EBfield4waves}), (\ref{eq:BEfield4waves}).
   As in the s-polarization case, in the limit of relatively low EM wave intensity  
   the electron performs the random walk motion comprised of short scale-length oscillations 
   interleaved by long scale excursions. An example of such trajectories is shown in 
   Fig.\,\ref{FIG-28} a) for the EM field amplitude $a_0=1.6\times 10^3$, 
 the dissipation parameter equal to $\varepsilon_{rad}=1.2\times10^{-8}$, 
 and the normalized critical QED field of $a_S=4.12\times 10^5$. The curve marked 
 by red color and the number ``1'' corresponds to the initial coordinates $x(0)=0.001$ and $y(0)=0.01$.
  Fig.\,\ref{FIG-28} b) presents a close-up of trajectory (1) the $(x,\,y)$ plane overlaid with the isocontours of the EM field ponderomotive potential. 
 Electron oscillations in the $(p_x,\,p_y)$ plane (Fig.\,\ref{FIG-28} c)) and dependence of  the $y$ coordinate on time plotted in   Fig.\,\ref{FIG-28} d) 
 demonstrate that the particle motion 
 is irregular. The stochastic character of the particle dynamics is also distinctly seen in the Poinca\'re sections in the plane  $(p_x,\,p_y)$, which is 
 presented in Fig.\,\ref{FIG-28} d). 

\begin{figure*}
	   \begin{center}
    \includegraphics[width=12 cm]{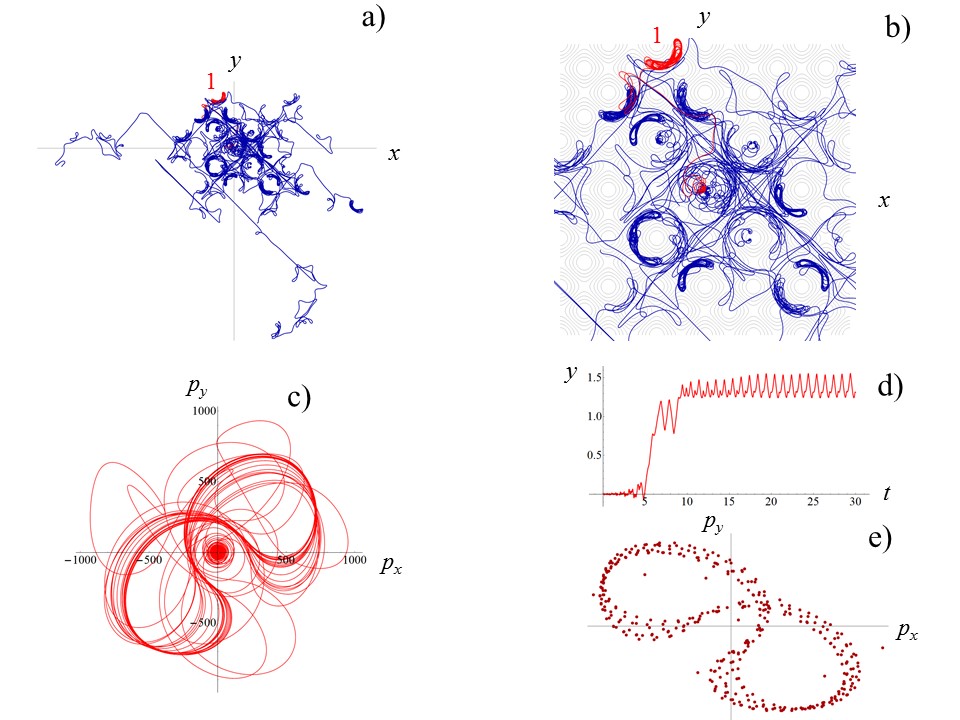}
		       \end{center}
 \caption{The same as in Fig.\,\ref{FIG-28} for $a_0=1.6\times 10^4$. 
  \label{FIG-29}}
 \end{figure*}

  For ten times higher EM field amplitude, when $a_0=1.6\times 10^4$,  the particle motion becomes regular as 
  seen in Fig.\,\ref{FIG-29}. In the $(x,\,y)$ plane  the electron performs long range L\'evy-like-flights along the lines
  $x=\pm y+\pm \pi n$, which end up in the localized attractors, where the particle oscillates pretty regularly (see Figs.\,\ref{FIG-29} a) and b)) . 
  This electron behavior is well seen in Figs.\,\ref{FIG-29} c) -- e) presenting the electron trajectory in the $(p_x,\,p_y)$ plane, the time 
  dependence of the $y$ coordinate and the Poinca\'re mapping in the momentum plane $(p_x,\,p_y)$, respectively. Broadening of the 
  trajectories in the Poinca\'re mapping  Figs.\,\ref{FIG-29} e) also indicates stochastic properties present in the particle motion.

\begin{figure*}
	   \begin{center}
    \includegraphics[width=12 cm]{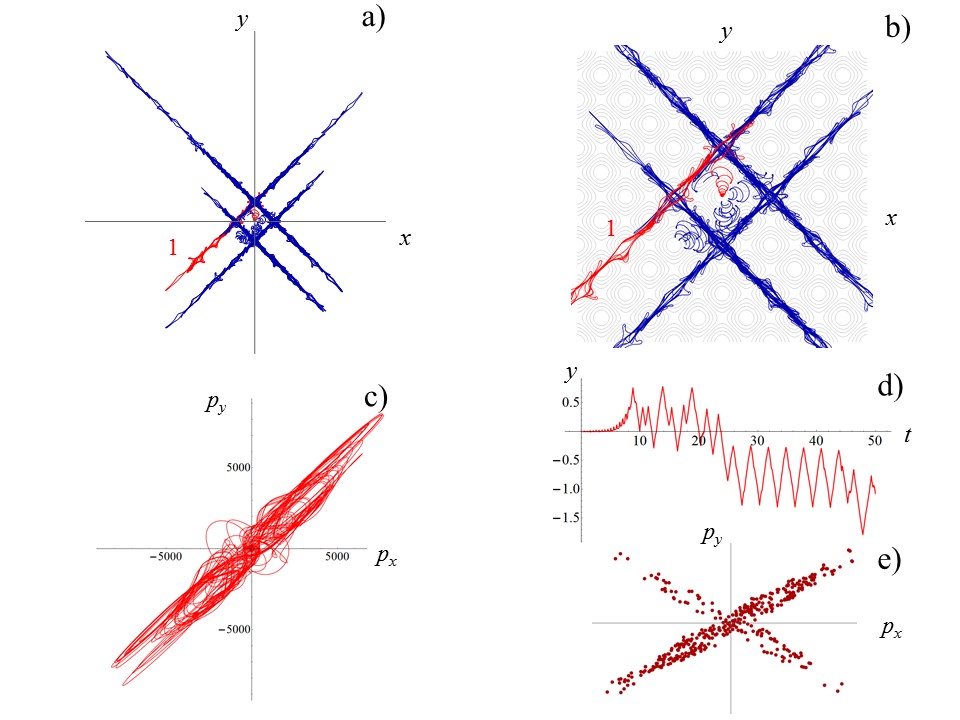}
		       \end{center}
 \caption{The same as in Fig.\,\ref{FIG-28} for $a_0=1.6\times 10^5$. 
  \label{FIG-30}}
 \end{figure*}

Further ten times increase of the EM field amplitude, $a_0=1.6\times 10^5$, results in the 
particle trapping within narrow stripes localized at the bottoms of the ponderomotive potential 
(Figs.\,\ref{FIG-30} a) and b)). A combination of regular and stochastic aspects of the particle 
dynamics in this case too is seen from the behavior of  the electron trajectory in the $(p_x,\,p_y)$ plane (Figs.\,\ref{FIG-30} c) ), from the time 
  dependence of the $y$ coordinate  (Figs.\,\ref{FIG-30} d) ), and from the broadening of the 
  trajectories in the Poinca\'re mapping  (Figs.\,\ref{FIG-30} d) ).

\section{Conclusions}

As is well known, the multiple colliding laser pulse concept \cite{SSB-2010a} is beneficial  for achieving extremely high amplitude of coherent
electromagnetic field (see also Refs. \cite{SSB-2010b, Gonoskov-2012, Gonoskov-2013, Gelfer-2015}). The complexity of the topology of the time-dependent EM field of colliding laser pulses results in the high complexity of the trajectories of charged particles interacting with these fields.
In the high field limit, when the radiation friction effects become significant, 
the charged particle behavior  demonstrates 
remarkable features corresponding to random walk trajectories, L\'evy flights, limit circles, attractors, and regular patterns. 

In the limit of the relatively weak laser intensity, the electron motion can be described as a random walk Fig. \ref{FIG-3bis} with the particle over-leaping from one field period  to another. The over-leaping correspond to the L\'evy flights. In contrast to the  standard theory of L\'evy flights, which can be found in Ref. \cite{LEVYFLIGHTS, LEVYFLIGHTS1, LEVYFLIGHTS2, LEVYFLIGHTS3}, in the 3 and 4 colliding waves case considered in the 
present paper, the L\'evy-like flights occur along the directions determined by the landscape of the ponderomotive potential determined in its turn by 
the geometry of the EM field of the colliding waves. Typically the particle performs short space scale (high frequency) oscillations intermittent 
with the long range leaps. This oscillation frequency appears to be significantly higher than the frequency of the driver EM wave due to the 
nonlinearity of the radiation friction force (see also discussion in Refs. \cite{Esirkepov-2015, Jirka-2016}). The length of the long range flight 
can be found from consideration of the charged particle momentum losses due to radiation friction as in Ref. \cite{NIMA-2011}.

Under certain conditions (in the high intensity and/or low frequency limit) the nonlinear dissipation mechanism  stabilizes the  particle motion causing the particle trapping within a narrow region located near the electric field maximum.
 In high intensity limit the particle can be trapped in the vicinity of the EM field ponderomotive potential performing 
regular motion there. The particle trajectory makes regular patterns shown in Figs. \,\ref{FIG-15}  and \,\ref{FIG-26}. 

We have analyzed the underlying physical mechanism of the radiating charge particle trapping in the regions of the electric field maximum. 
As elucidated within the framework of the simple model formulated in the present paper  the particle trapping is explained by the friction drag 
originating from the nonlinear dependence of the radiation friction on the EM field. 

The attractor trajectory patterns in Figs.\,\ref{FIG-15}, \ref{FIG-17} and \,\ref{FIG-25} are made by an ensemble of electrons. 
The single electron trajectory shown in Fig.\,\ref{FIG-16} demonstrates that having been moving 
for a long enough time it could cover the whole attractor. In view of this, there are two questions.
The first one being is there an analogy of the ergodic hypothesis saying that over long periods of time, the time spent 
in some region of the attractor is proportional to the attractor measure? The second one being is there an analogy of the 
Poincar\'e recurrence theorem (\cite{Arnold-1989}) saying that the particle, after a sufficiently long but finite time, returns to a point very close 
to the initial point?  A similar question occurs in the case of the particle random walk on whether 
the results of the well known random walk theory (see \cite{MKac-1947}) 
can be used in our case.
Since finding the answers to these questions requires additional thorough consideration, we leave this to the forthcoming publications. 

One of the most important findings of the present work is a revealing of  a new class of regular distributions made by ensembles of the particle trajectories. They are structurally determinate  patterns, as if made by tiles, formed in the high field amplitude limit when the radiation friction force drastically modifies the charged particle dynamics in the electromagnetic field as can be distinctly seen in Figs. \ref{FIG-16},  \ref{FIG-18},  \ref{FIG-25}, and  \ref{FIG-26}.  As for the  possible practical implications of these findings, these ``crystal-like'' 
patterns are expected to be seen in the spatial distribution of the gamma-rays  emitted by the electrons irradiated by the multiple high power laser 
pulses, which has been noticed in Refs. \cite{LS-2016} and \cite{Y-2016}.

\section*{Acknowledgments}

SSB acknowledges support from the Office of Science of the US DOE under Contract No. DE-AC02-05CH11231.
XQY and ZG acknowledge support from National Basic Research Program of China (Grant No.2013CBA01502), NSFC (Grant Nos.11535001) and National Grand Instrument Project (2012YQ030142).



%

\end{document}